\documentclass[twocolumn,         
               showkeys,showpacs,            
               preprintnumbers,     
               aps,                 
               prd,longbibliography,          	    
               letterpaper,             
               superscriptaddress,      
               nofootinbib,         
               tightenlines,        
               floats,floatfix      
               ]{revtex4-1}
\usepackage{dcolumn}   
\usepackage{bm}        
\usepackage{float}
\usepackage{physics}
\usepackage{graphicx,color}
\usepackage{latexsym}
\usepackage{amsmath,amssymb}        
\usepackage[colorlinks=true,linkcolor=blue,citecolor=blue]{hyperref}
\usepackage{mathrsfs}
\usepackage{comment}
\usepackage{soul}
\usepackage{cancel}
\definecolor{purple}{rgb}{0.58,0.0,0.83}
\definecolor{orange}{rgb}{1,0.5,0}
\DeclareSymbolFontAlphabet{\mathrsfs}{rsfs}
\DeclareMathAlphabet{\mathcal}{OMS}{cmsy}{m}{n}

\begin{document}


\title{Possible formation mechanism of multistate gravitational atoms}


\author{F. S Guzm\'an}
\email{francisco.s.guzman@umich.mx}
\affiliation{Instituto de F\'{\i}sica y Matem\'{a}ticas, Universidad
              Michoacana de San Nicol\'as de Hidalgo. Edificio C-3, Cd.
              Universitaria, 58040 Morelia, Michoac\'{a}n,
              M\'{e}xico.}


\date{\today}


\begin{abstract}
The collision of two equilibrium ground state solutions of the Schr\"odinger-Poisson (SP) system, in orthogonal states, is proposed as a formation mechanism of mixed state solutions of the SP system with spherical and first dipolar components. The collisions are simulated by solving numerically the SP system for two orthogonal states, considering head-on encounters, and using various mass ratios between the initial configurations with different head-on momentum. The results indicate that the less massive of the configurations pinches off the more massive one, and redistributes its density along the axis of collision. The averaged in time density of the two states resembles the distribution of matter of bistate equilibrium configurations with monopolar and dipolar contributions. 
\end{abstract}




\maketitle

\section{Introduction}

Boson stars (BSs), stationary solutions of self-gravitating scalar fields, have received attention recently due to important state of the art astrophysical scenarios. On the relativistic regime, BSs have been candidates as mimickers of supermassive black holes \cite{Torres2000,Torres2002,Narayan2004,Guzman2006} and currently the idea is contrasted with state of the art observations \cite{Olivares2020}. On the Newtonian regime, since ultralight bosonic dark matter \cite{Matos:2000ss,Sahni:2000} has grown a strong dark matter candidate \cite{Hui:2016}, Newtonain boson stars play the role of attractor clumps of dark matter in structure formation simulations \cite{Schive:2014dra,Mocz:2017wlg}, commonly called solitonic density profiles. In these two scenarios, relevant boson stars correspond to the ground state configurations as described originally in \cite{Ruffini:1969}. 

Excited state configurations have been considered since the very beginning of the study of these objects \cite{Ruffini:1969}, whereas clear results on the instability of excited states toward ground state configurations in the relativistic \cite{SeidelSuen1998} and Newtonian regimes \cite{GuzmanUrena2006} have been demonstrated with numerical simulations. Multistate configurations on the other hand, have recently been developed also within the relativistic \cite{SanchisGual} and Newtonian \cite{BernalUrena2010,GuzmanUrena2020} regimes, so as initial stability analyses \cite{SanchisGual,Guzman2021}.

The applicability of multistate configurations within astrophysics is still unforeseen, because single state scenarios are the ones mostly considered. Theoretical analyses include the study of binary configurations of spherical multistate solutions, during collisions \cite{GuzmanAvilez2018} and their relaxation processes \cite{GuzmanAvilez2019}.
As far as we can tell, there is at least one foreseeable application, related to the vast polar structure (VPOS) problem of satellite galaxies whose orbital distribution around host major galaxies is not isotropic as predicted by cold dark matter (CDM) simulations \cite{Bullock1,Bullock2}; instead they are distributed on preferential planes  \cite{VPOSNature,VPOSConn_2013,VPOS1,VPOS2}. The application is within the ultralight bosonic dark matter, and assumes that galactic halos of host galaxies can be bistate equilibrium configurations of bosons consisting of combinations of ground (1,0,0) and first dipolar contribution (2,1,0) states \cite{jordi2021}. The bistate galactic dark matter model is motivated by a number of preliminary and simple consistency checks, including that Milky Way rotation curve is well explained, that bistate configurations are long-lived, and that galactic disks of test particles are maintained and survive the effects of the dipolar gravitational potential contribution as long as the disk is not tilted with respect to the equatorial plane \cite{jordi2021}. Test particles around these configurations accommodate on planar orbits, with higher probability on planes passing near galactic poles at distances bigger than 30 kpc depending on the monopole to dipole mass ratio, and the results show consistency with observations that depends on the monopole to dipole mass ratio \cite{jordi2021}. Also differences in behavior of test particles around  similar triaxial CDM halos are clear and quantified. The main potential contribution of this model resides in the low probability that satellites accommodate on such inclined planes for the three well observed galaxies, namely the Milky Way, M31 and Centaurus A without a nonspherical contribution of galactic dark matter, whereas multistate solutions exist for bosonic dark matter.

These simple consistency checks are very basic, and more sophisticated models than test particles representing the behavior of dwarf galaxies are needed in order to restrict the strength of the model. Meanwhile another very important condition is needed, if multistate configurations will be considered in astrophysical scenarios it is necessary to know whether they can be formed. In this manuscript we present a possible formation mechanism of bistate configurations with spherical and first dipole components (1,0,0)+(2,1,0), based on numerical simulations. The idea is that the collision of  two equilibrium ground state solutions in orthogonal states can lead to the formation of configurations similar to the bistate ones. Beyond this incipient motivation, it is  interesting to study whether multistate configurations can ever be formed, and possible future applications may arise.

The paper is organized as follows. In Sec. \ref{sec:multi} we briefly review the construction and properties of bistate configurations, in Sec. \ref{sec:model} we illustrate the formation model, in Sec. \ref{sec:results} we show the results and analysis of our simulations and in Sec. \ref{sec:conclusions} we draw some conclusions.

\section{Multi-state configurations}
\label{sec:multi}

Before explaining the formation process we briefly review what bistate configurations are. The equations that rule multistate configurations of the SP system are

\begin{eqnarray}
i\frac{\partial \Psi_{nlm}}{\partial t} &=& -\frac{1}{2}\nabla^2 \Psi_{nlm} + V\Psi_{nlm},\label{eq:schro}\\
\nabla^2 V &=& \sum_{nlm} |\Psi_{nlm}|^2,\label{eq:Poisson}
\end{eqnarray}

\noindent where $\Psi_{nlm}$ is the wave function of the state $nlm$ and $V$ is the gravitational potential coupling the various states. Notice that Poisson equation does not contain the crossed terms among different states, because each state is assumed to have a definite number of particles and there is no mixture. Also, when the SP system is considered the low energy and weak field limit of the Einstein-Klein-Gordon system, the normal ordering condition requiring  vacuum to have a zero density  energy density, implies all the crossed terms vanish \cite{BernalUrena2010}.

As described in \cite{GuzmanUrena2020}, these equations have stationary solutions, constructed based on the assumption of harmonic time dependence for each state $nlm$

\begin{equation}
\Psi_{nlm}(t,{\bf x}) = \sqrt{4\pi} e^{-i\gamma_{nlm}}r^l \psi_{nlm}(r)Y_{lm}(\theta,\phi),
\label{eq:ansatz}
\end{equation}

\noindent where each radius-dependent wave function $\psi_{nlm}(r)$ has the associated eigenfrequency $\gamma_{nlm}$. This harmonic behavior guarantees that $|\Psi_{nlm}|^2$, and consequently the gravitational potential is time-independent. A stationary Schr\"odinger equation is defined for each state $\psi_{nlm}$ that is solved numerically as an eigenvalue problem provided isolation boundary conditions and regularity at the origin for the wave function of each state \cite{GuzmanUrena2020}.

For the case of a configuration with states $(1,0,0)$ and $(2,1,0)$, two parameters label the family of solutions of the eigenvalue problem, which are the central value of the stationary wave functions $\psi_{100}(0)$ and $\psi_{210}(0)$. By fixing $\psi_{100}(0)=1$ the contribution of the dipolar component is parametrized only by the central value $\psi_{210}(0)$.

The mass ratio between the two states is an important parameter that determines the contribution of the dipolar state to the gravitational potential of the configuration. Defining the mass of states $(1,0,0)$ and $(2,1,0)$ as
$M_{100}=\int |\Psi_{100}|^2d^3 x$ and $M_{210}=\int |\Psi_{210}|^2d^3 x$ we define the dipole to monopole mass ratio as $MR=M_{210}/M_{100}$.

\begin{figure}[htp]
\includegraphics[width=3.5cm]{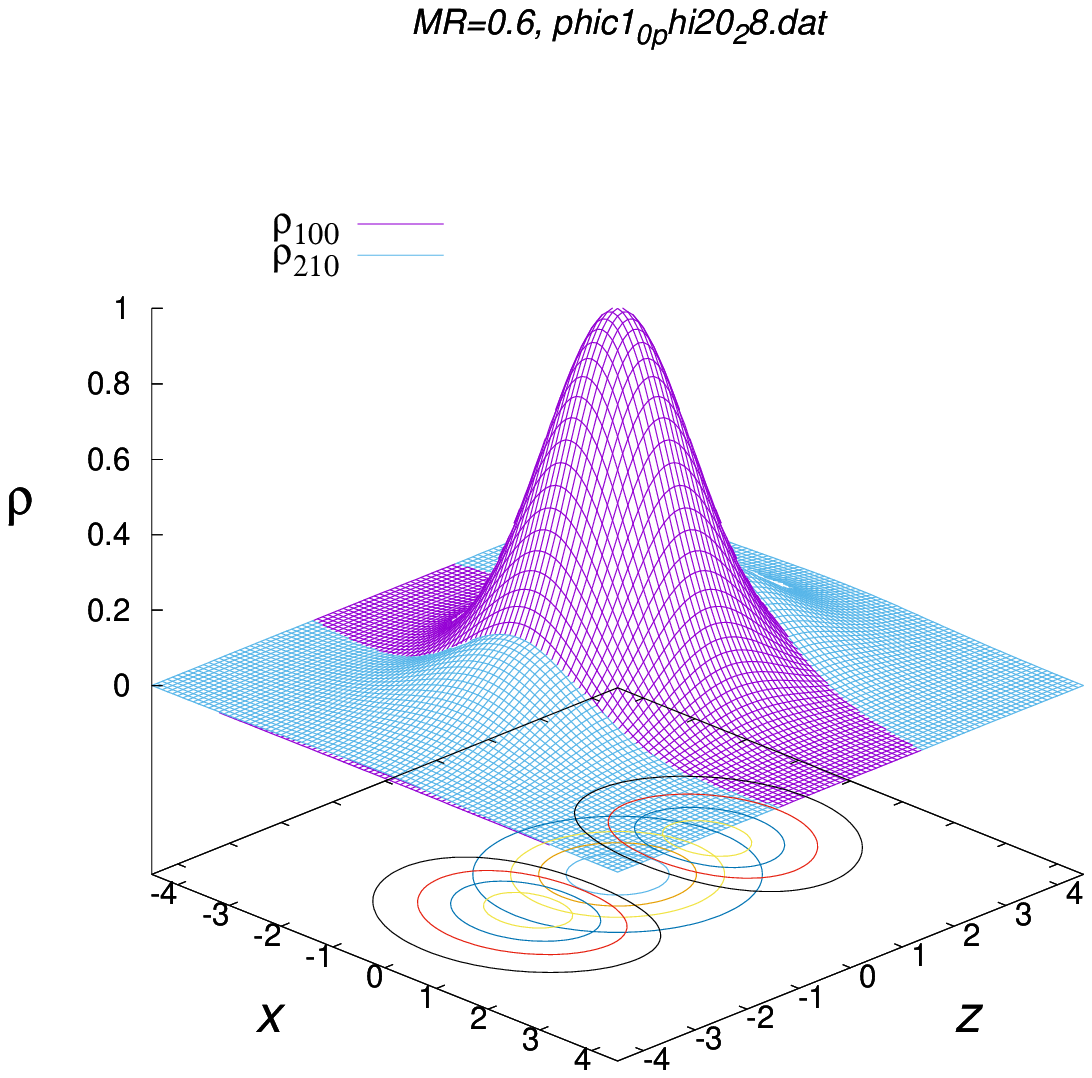} 
\includegraphics[width=3.5cm]{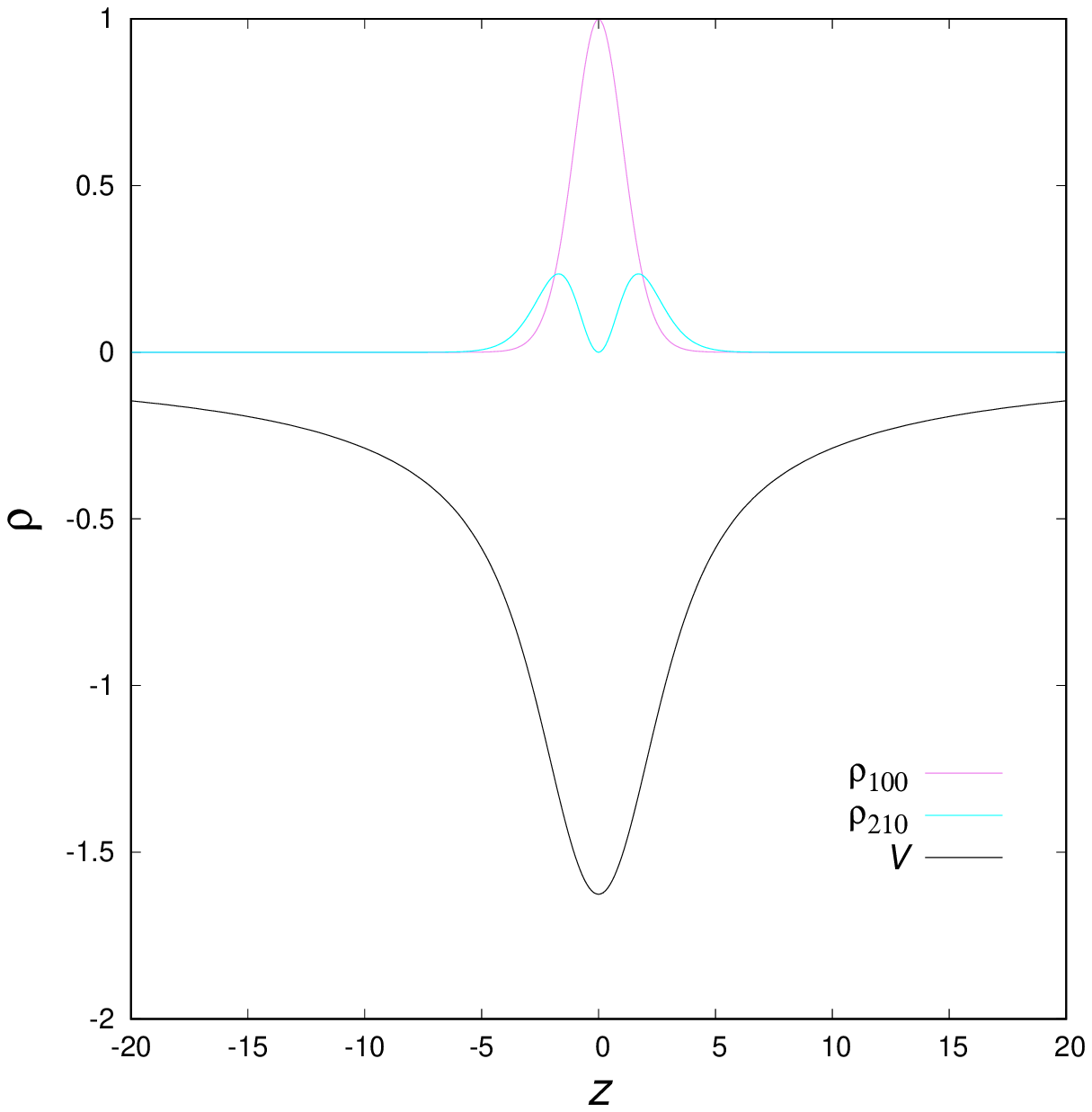} \\
\includegraphics[width=3.5cm]{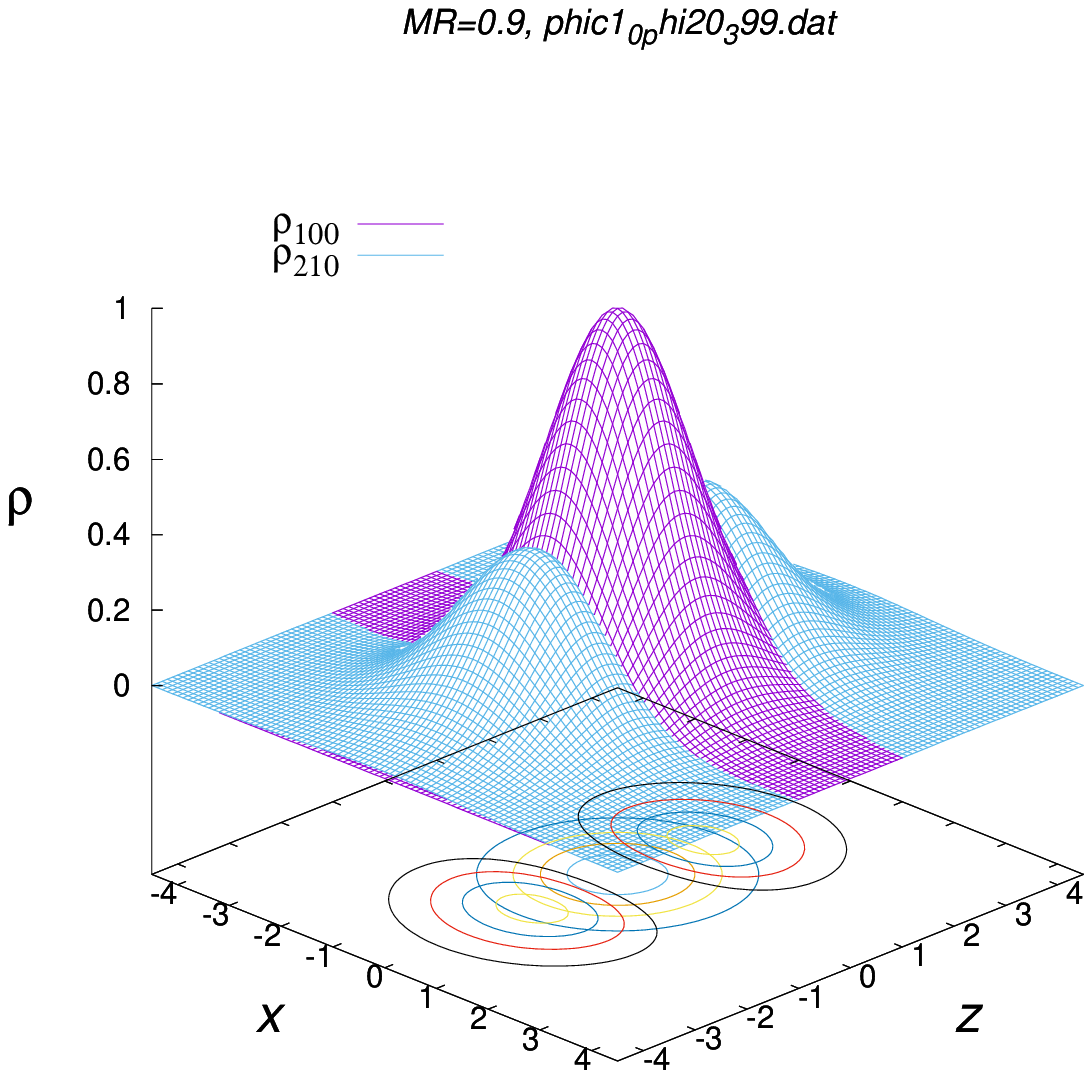} 
\includegraphics[width=3.5cm]{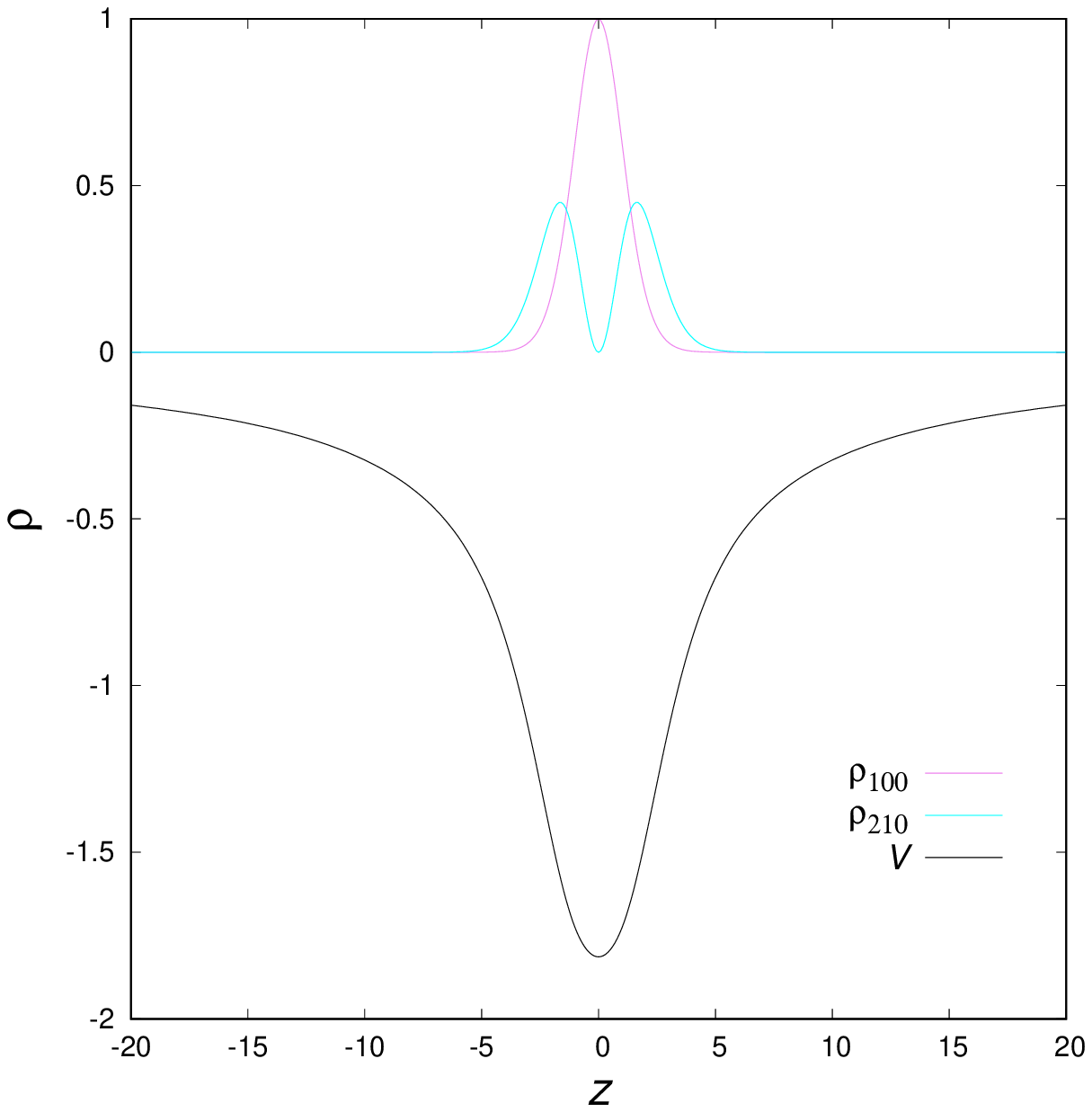} 
\includegraphics[width=3.5cm]{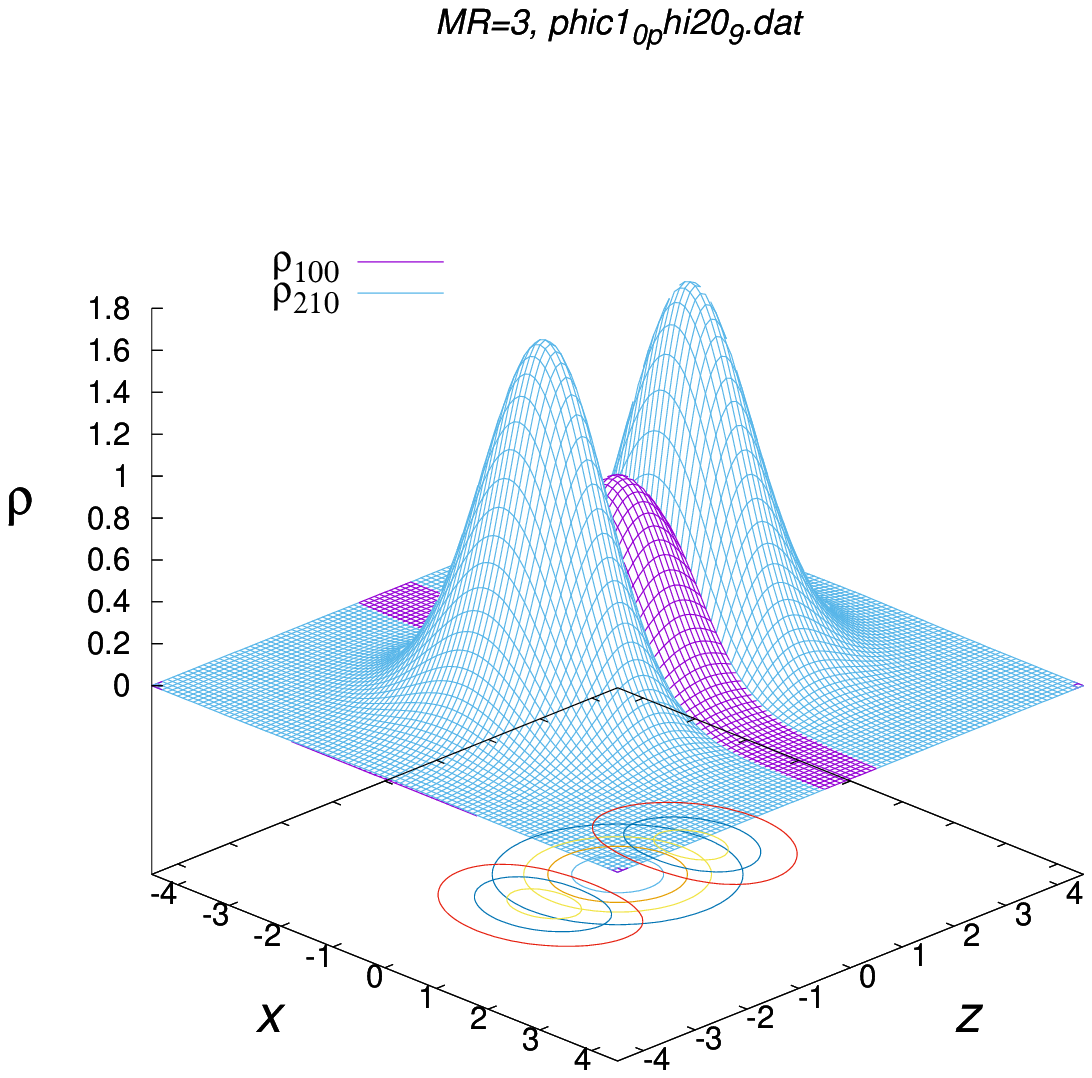} 
\includegraphics[width=3.5cm]{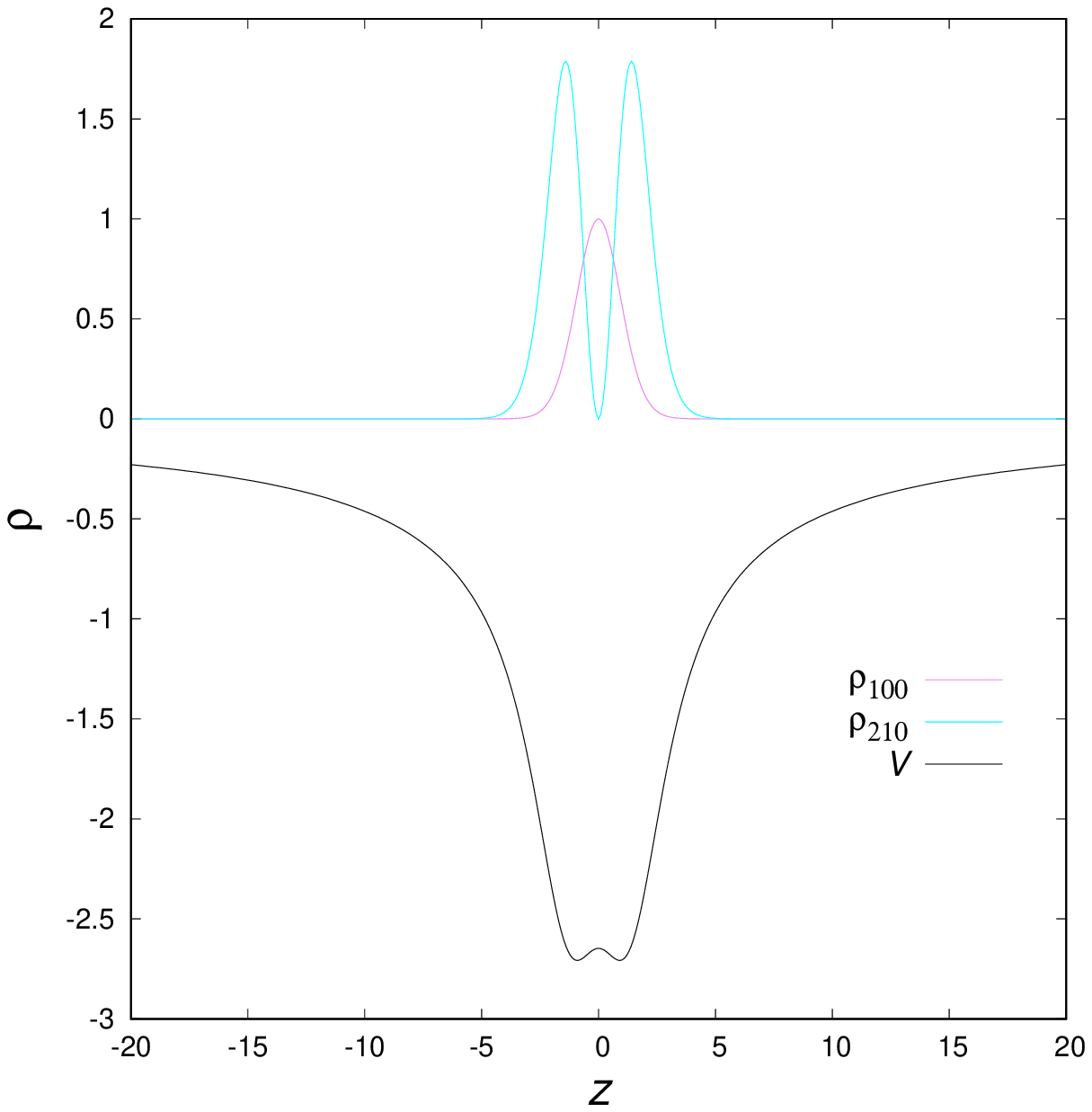} 
\caption{\label{fig:equilibrium} (Left) Densities $\rho_{100}$ and $\rho_{210}$ projected on the $xz-$plane. The plots correspond to configurations with $MR=0.6$, $MR=0.9$ and $MR=3$. The first two are monopole dominated configurations and are of the type of configurations that can possibly be formed according to our results, whereas the third configuration is dipole dominated and is not found to form in our analysis. (Right) Projection of the densities along the $z-$axis together with their corresponding gravitational potential $V$.}
\end{figure}

In Fig. \ref{fig:equilibrium} we show three configurations with different mass ratio $MR$. The first two correspond to a monopole dominated configuration with $MR<1$ whereas the third one is dominated by the dipole with $MR>1$. The dipole dominance in the third case is also reflected in the double bump shape of the gravitational potential along the $z-$axis.

\section{Model of formation}
\label{sec:model}

Equilibrium stationary ground state configurations are of considerable interest because they are stable and attractors in time, locally \cite{GuzmanUrena2006,BernalGuzman2006b} and from structure formation simulations \cite{Schive:2014dra,Mocz:2017wlg}. It is expected that these configurations form and then evolve according to their own gravity and the local environment.

Bosonic dark matter models assume bosons are in a single coherent state, not in two or more. The basic assumption in our model is that equilibrium configurations can be formed in different states and can collide. The merger of two ground state, spherically symmetric equilibrium configurations, in orthogonal states $\Psi_1$ and $\Psi_2$, obey the system

\begin{eqnarray}
i\frac{\partial \Psi_{1}}{\partial t} &=& -\frac{1}{2}\nabla^2 \Psi_{1} + V\Psi_{1},\label{eq:schro1}\\
i\frac{\partial \Psi_{2}}{\partial t} &=& -\frac{1}{2}\nabla^2 \Psi_{2} + V\Psi_{2},\label{eq:schro1}\\
\nabla^2 V &=& |\Psi_{1}|^2 + |\Psi_{2}|^2,\label{eq:Poisson12}
\end{eqnarray}

\noindent which is the evolution system of equations (\ref{eq:schro})-(\ref{eq:Poisson}) for two states. 

The spherically symmetric ground state configurations to be collided, are constructed following the recipe in \cite{GuzmanUrena2004} using spherical coordinates. The initial wave functions for $\Psi_1$ and $\Psi_2$ are associated with densities $\rho_1=|\Psi_1|^2$ and $\rho_2=|\Psi_2|^2$. We choose configuration 1 to have central density $\rho_{1c}=1$ with mass in numerical units $M_1=2.602$ and radius $R_{1,95}\sim3.93$, containing 95\% of the total mass according to \cite{GuzmanUrena2004}. For configuration 2 we set the mass to $M_2=\lambda M_1$, radius $R_{2,95}\sim3.93/\lambda$, and wave function $\Psi_2=\lambda^2 \Psi_1$, where $\lambda$ is the initial mass ratio between the two configurations. This is done using the scaling relations $\{r,\Psi,M \}\rightarrow \{\hat{r}/\lambda,\lambda^2\hat{\Psi},\lambda \hat{M} \}$ for $\lambda$ any constant, in particular our mass ratio, if the hatted configuration is a ground state configuration the unhatted one is also an equilibrium configuration \cite{GuzmanUrena2004}.

We simulate the head-on collision of these two equilibrium configurations on a numerical domain described with  cylindrical coordinates $(r,\phi,z)$ as done in \cite{BernalGuzman2006a,BernalGuzman2006b}. We use the domain $D=[0,r_{max}]\times[z_{min},z_{max}]$ that we discretize with $N_r$ and $N_z$ squared cells. Numerical methods can be summarized as follows. We use the method of lines to evolve the wave functions in Schr\"odinger equations, which uses a third order Runge-Kutta integrator and right hand sides discretized with fourth order stencils. Poisson equation is solved with a successive over-relaxation method and multipolar boundary conditions at the boundary of the domain at each time step. A sponge type condition is used to prevent the wave function modes approaching the boundary to be reflected back into the numerical domain where the collision and further evolution takes place \cite{Grupo2014}.

Initial conditions for the head-on encounter consist in centering configuration 1 at $(0,0,\lambda z_0)$ and configuration 2 at $(0,0,- z_0)$ so that the center of mass is approximately at the origin of coordinates. The wave functions $\Psi_1$ and $\Psi_2$ are injected into the numerical domain $D$ using bilinear interpolation.

Head-on momentum $p_{z}>0$ is added to configuration 2 and $p_{1z}=-\lambda p_{z}$ to configuration 1, in order to keep the center of mass at the coordinate origin.
The momentum is added to the wave function of each configuration by rescaling $\Psi_1 \rightarrow e^{p_{1z}}\Psi_1$ and $\Psi_2 \rightarrow e^{p_{z}}\Psi_2$. After the head-on momentum is applied we solve Poisson equation (\ref{eq:Poisson12}) at initial time, which completes the initial condition of the problem.

We evolve the system in a domain with $r_{max}=z_{max}=-z_{min}=30$, discretized with $N_z=2N_r=300$ cells. The initial position of the configuration is $z_0=10$ and is such that $\langle \Psi_1,\Psi_2 \rangle < 10^{-10}$ in the whole domain at initial time. This resolution allows self-convergence of order between second and third.

\section{Parameter space}
\label{sec:results}

\begin{figure}[htp]
\includegraphics[width=2.5cm]{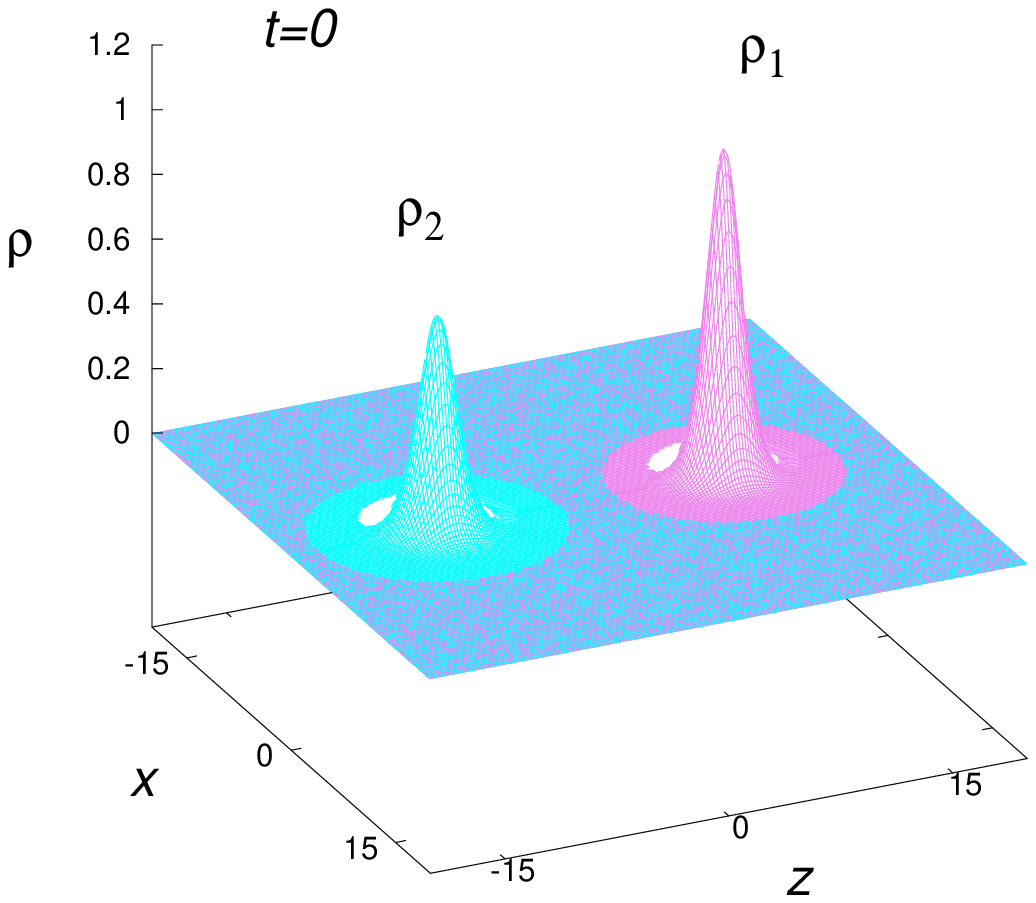} 
\includegraphics[width=2.5cm]{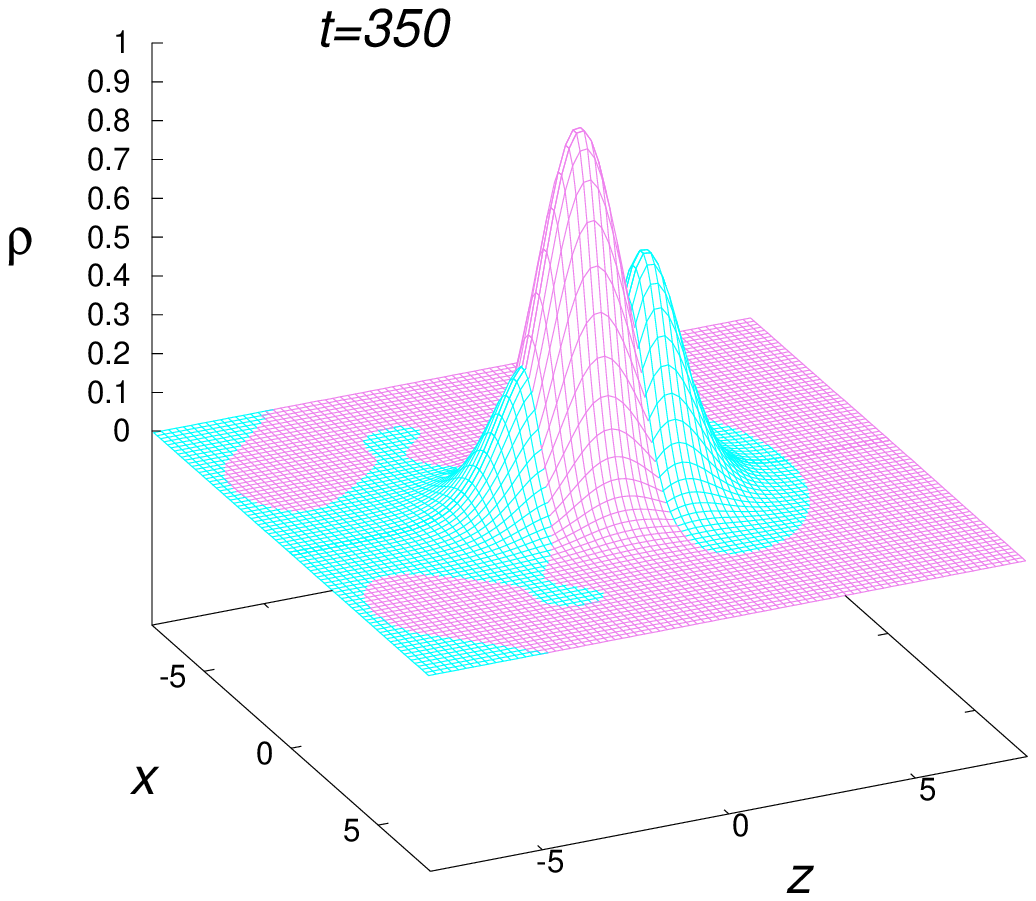} 
\includegraphics[width=2.5cm]{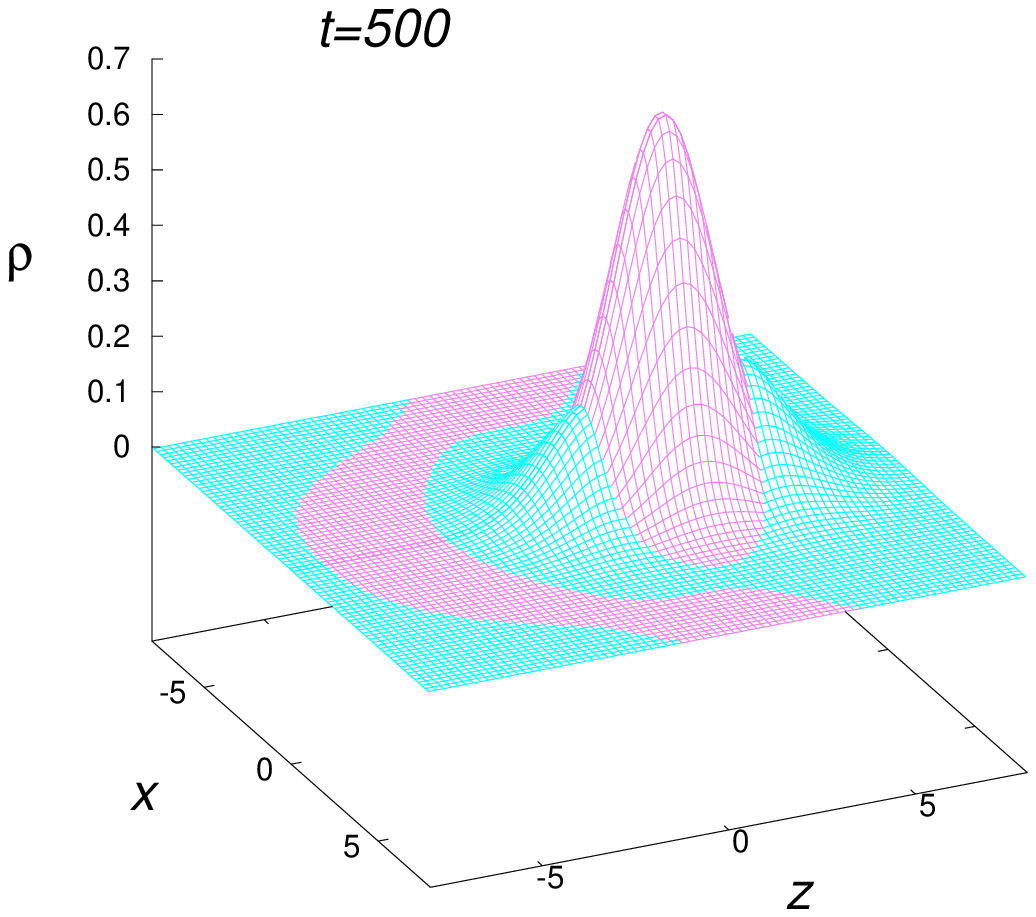} 
\includegraphics[width=2.5cm]{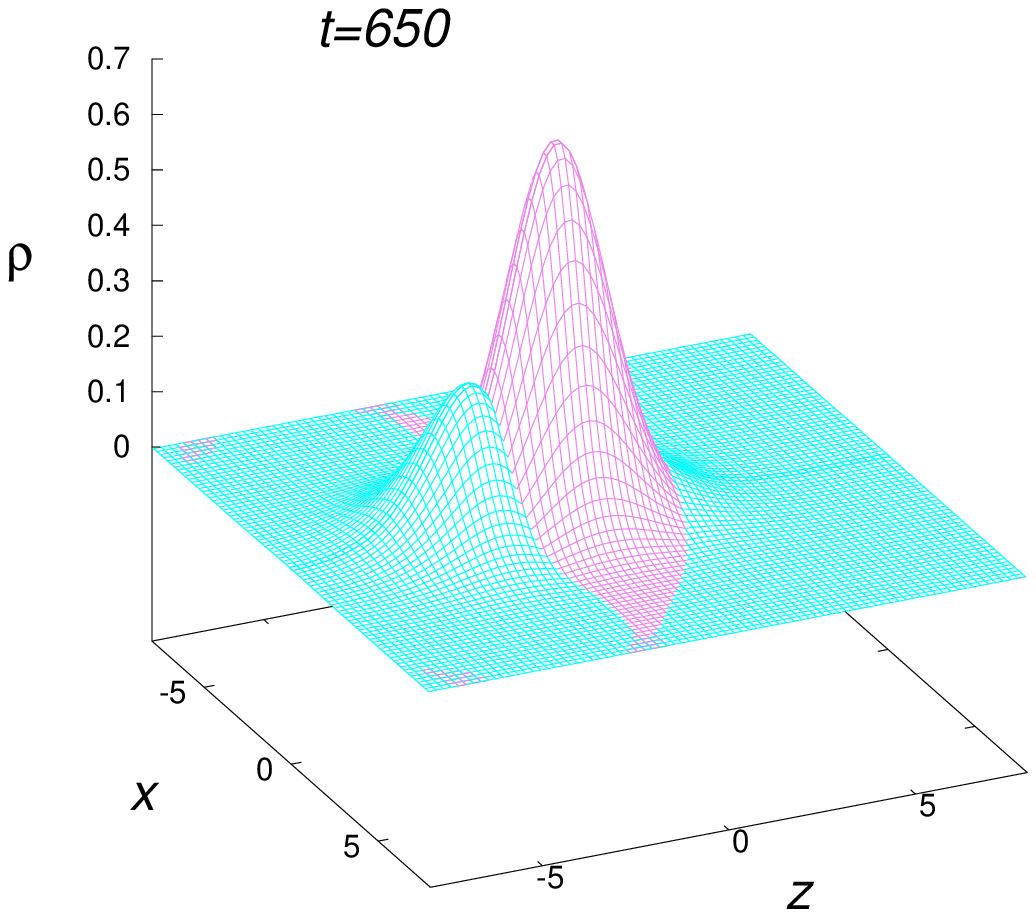} 
\includegraphics[width=2.5cm]{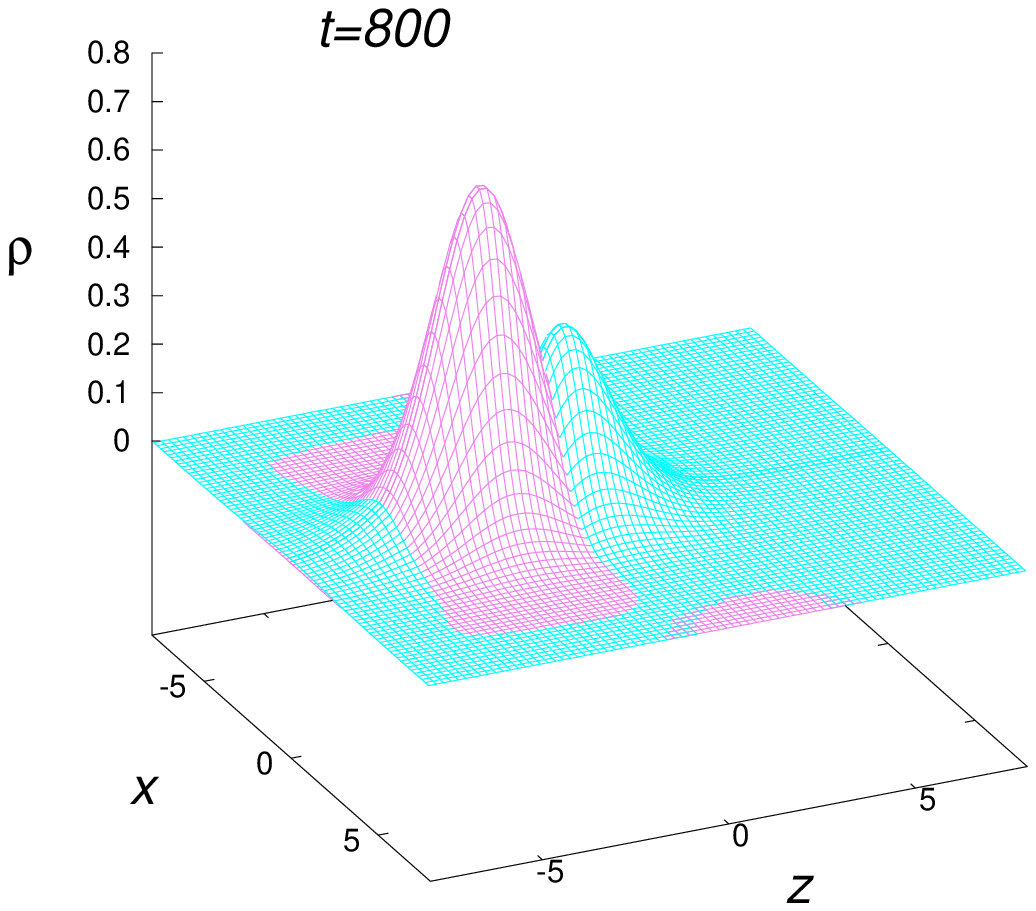} 
\includegraphics[width=2.5cm]{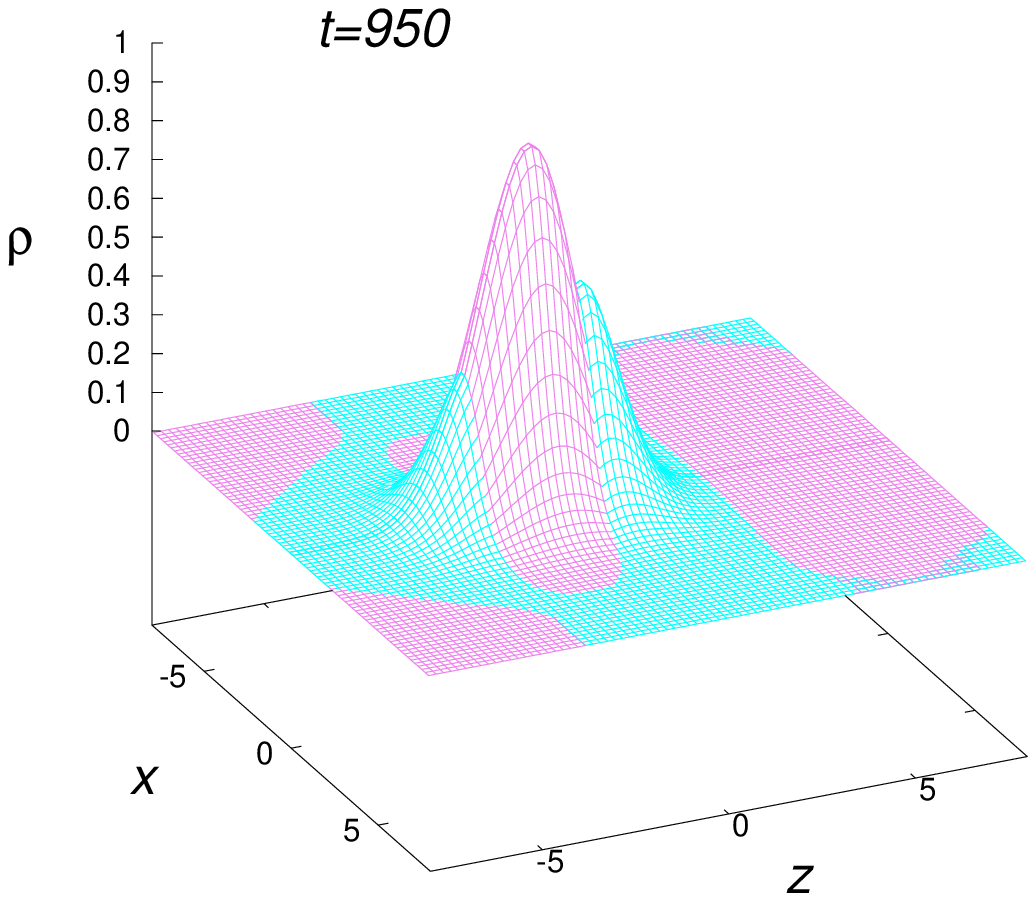} 
\caption{\label{fig:mr0_9_pz0_2_snaps} Snapshots of densities $\rho_1$ and $\rho_2$ projected on the $xz-$plane for the case with parameters $MR=0.9$ and $p_z=0.2$. In the plots we extended the $rz-$domain to the cartesian $xz-$ plane for the sake of illustration. The first snapshot is at initial time and the following ones cover approximately a period of the oscillation around the origin.}
\end{figure}

We explore the parameter space of initial mass ratio $MR=0.5,0.7,0.9$ and various values of the head-on momentum $p_z$. 
In Fig. \ref{fig:mr0_9_pz0_2_snaps} we illustrate the generic behavior of the mergers with the case $MR=0.9$ and $p_{z}=0.2$. The whole system itself oscillates around the coordinate origin, and the blobs of density $\rho_2$ change amplitude and evolve aside the blob with density $\rho_1$. The evolution of various collisions can be seen in the animations\footnote{https://sites.google.com/umich.mx/fsguzman/supplemental-materials}.

In all cases the results are similar; namely, the small mass configuration pinches off the bigger configuration and its density $\rho_2$ splits into two blobs, one at each side of the density $\rho_1$.

\begin{figure}[htp]
\includegraphics[width=2.5cm]{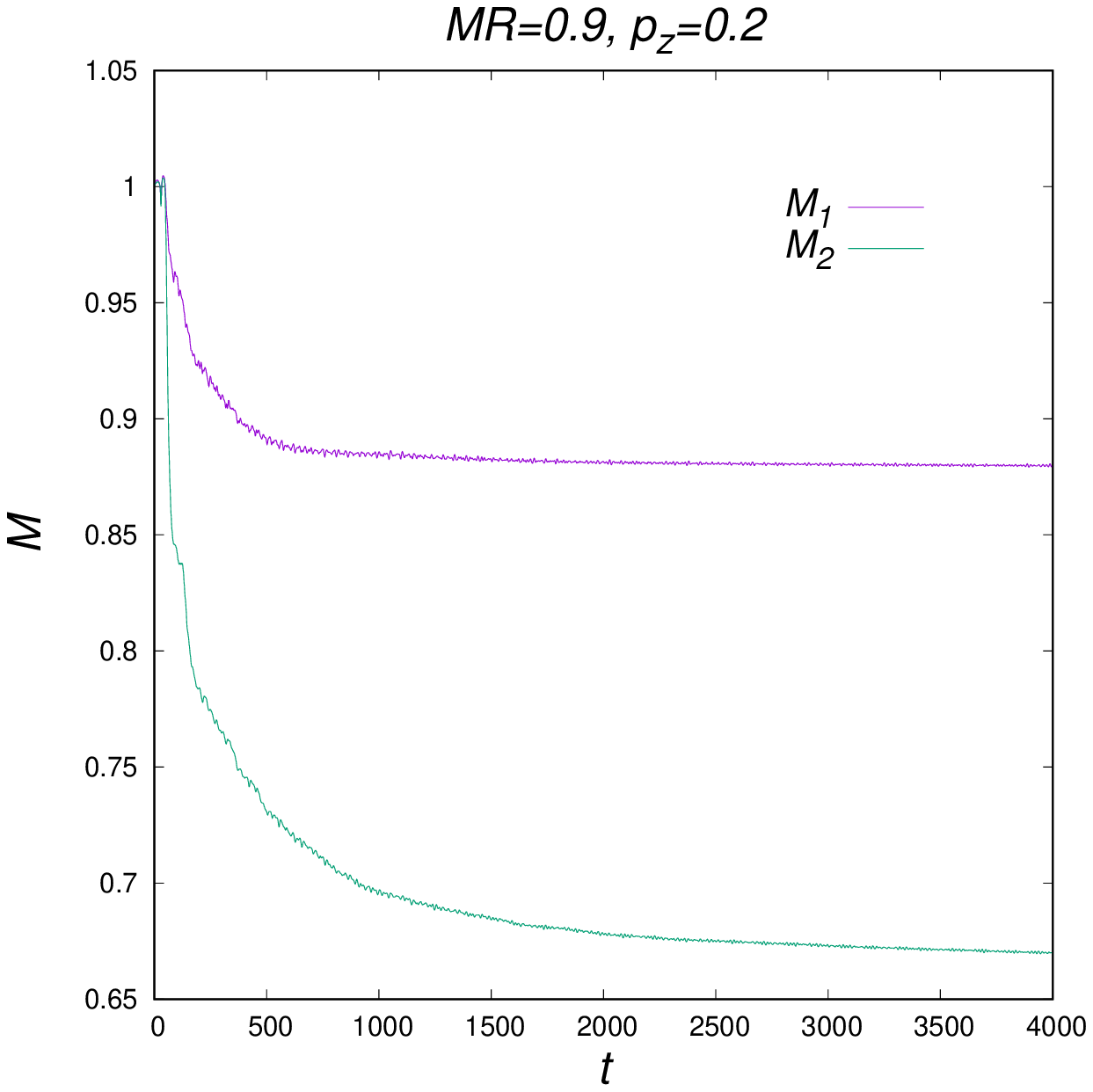} 
\includegraphics[width=2.5cm]{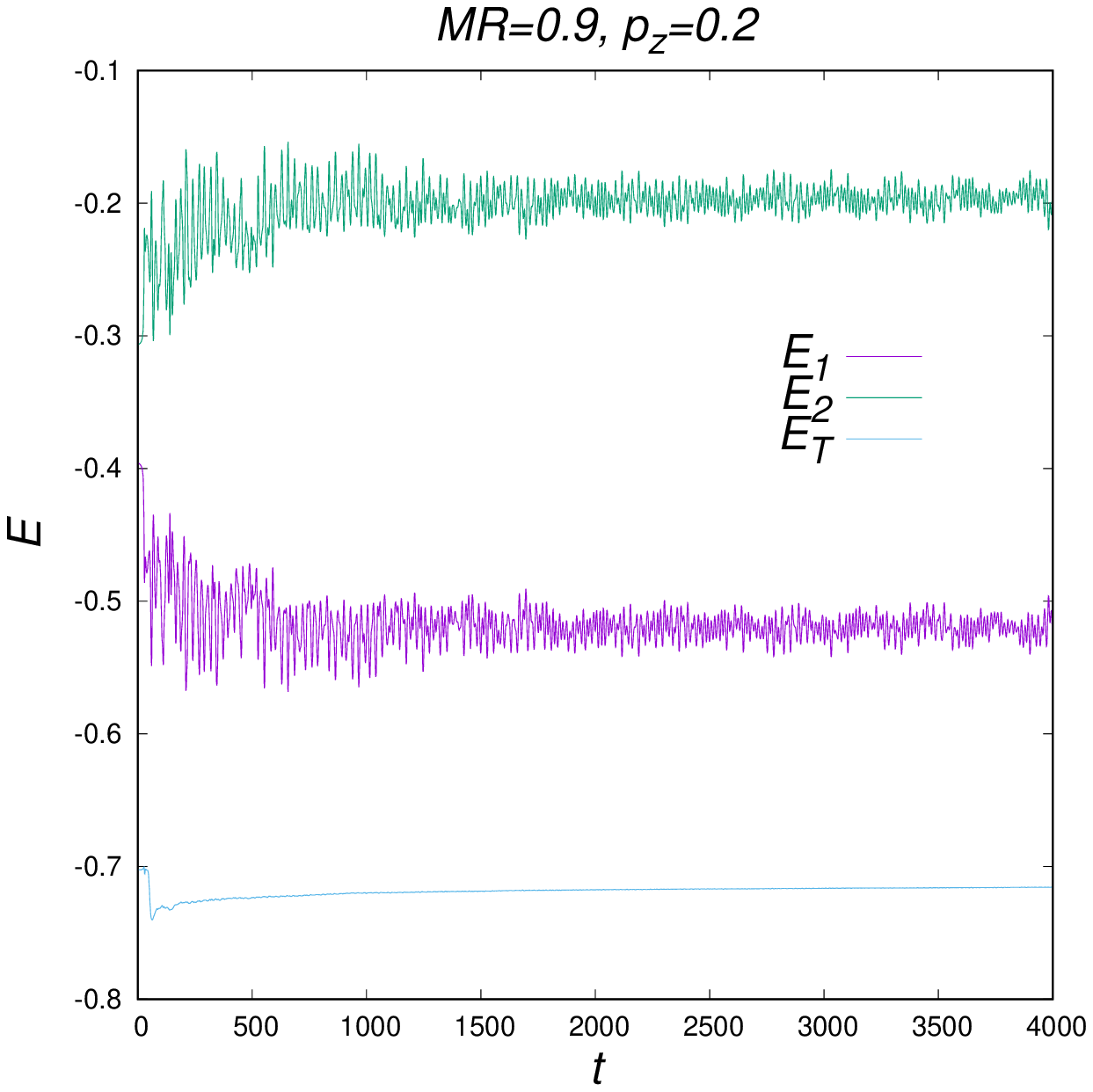} 
\includegraphics[width=2.5cm]{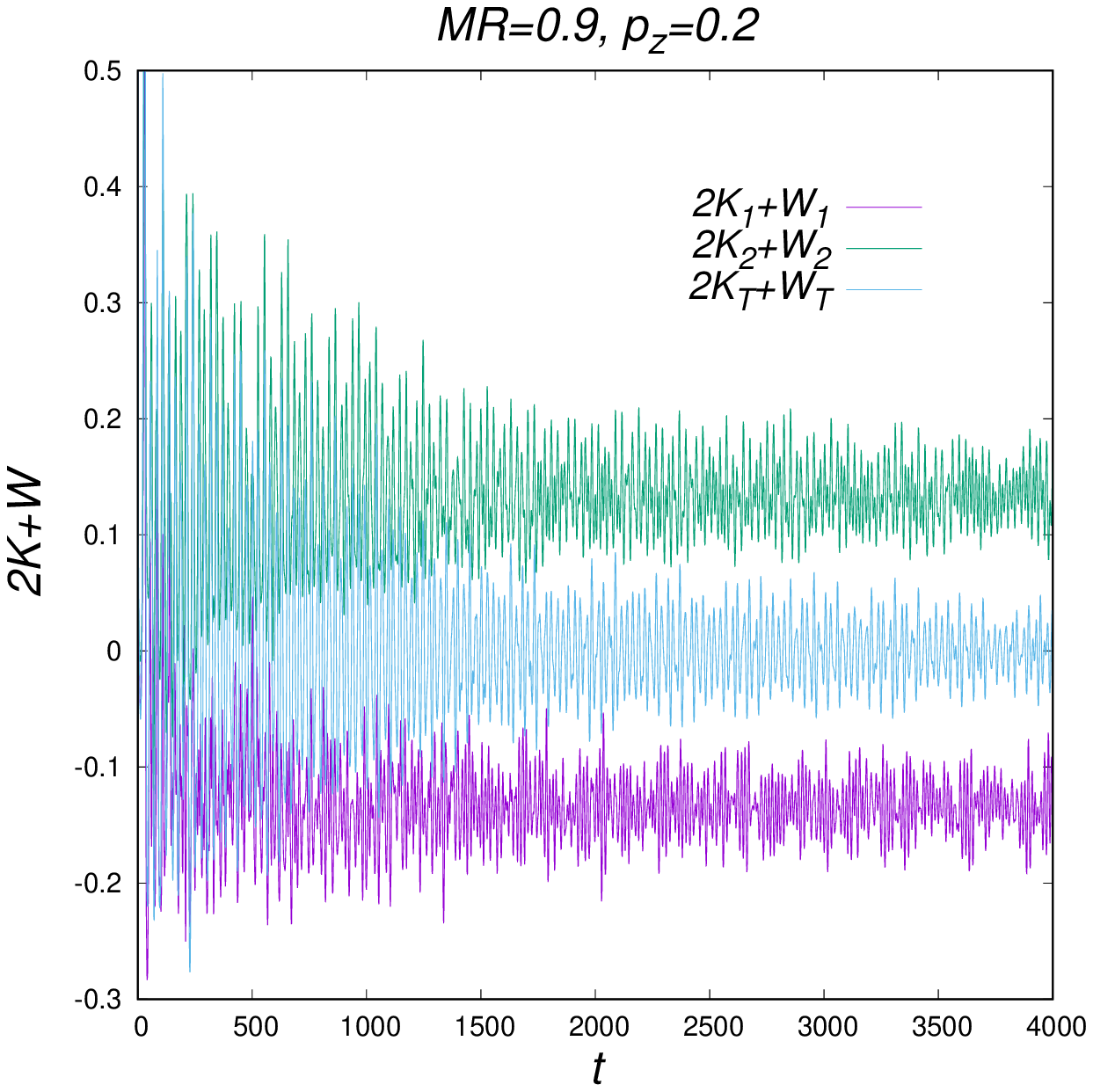} 
\includegraphics[width=2.5cm]{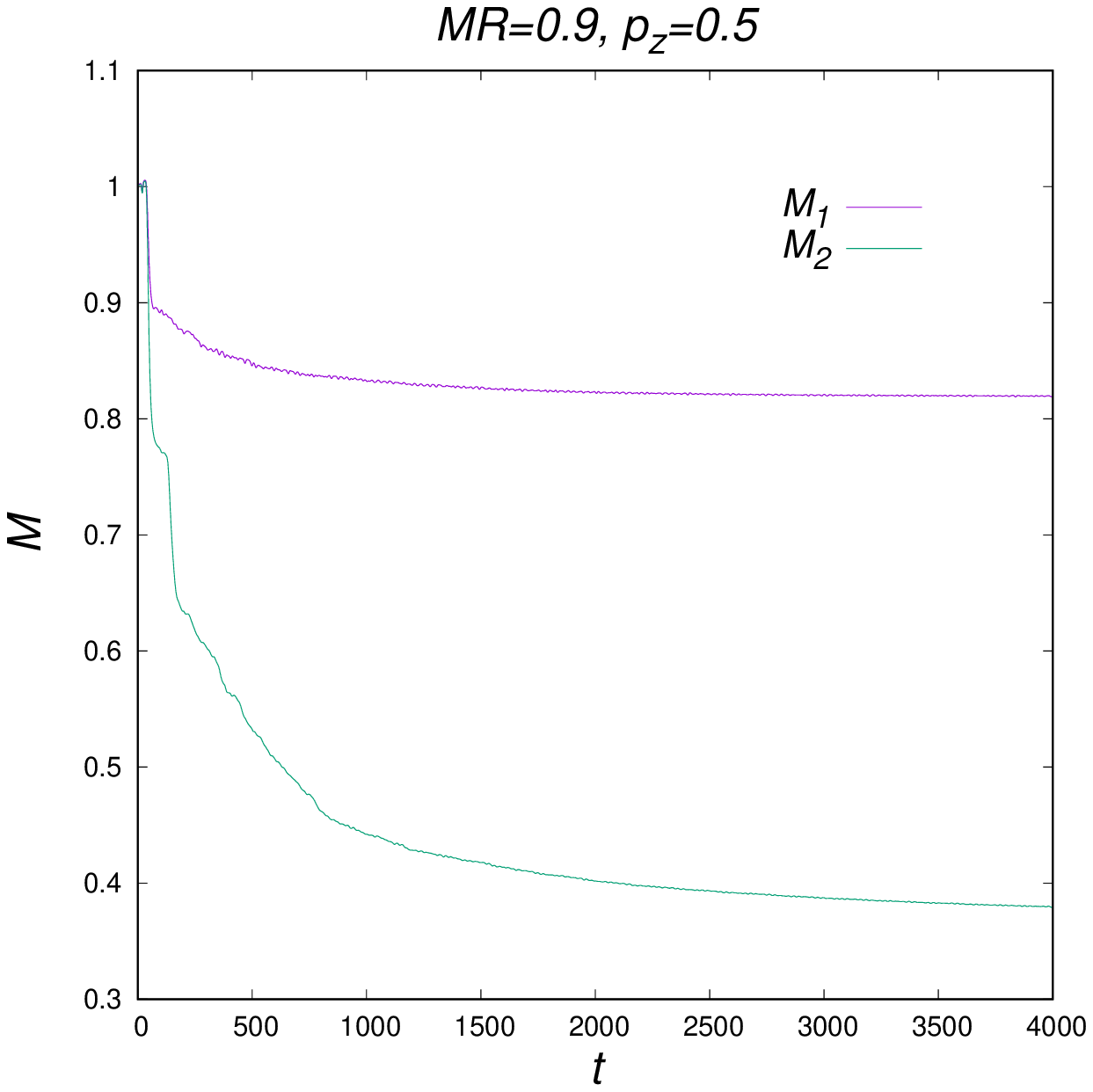} 
\includegraphics[width=2.5cm]{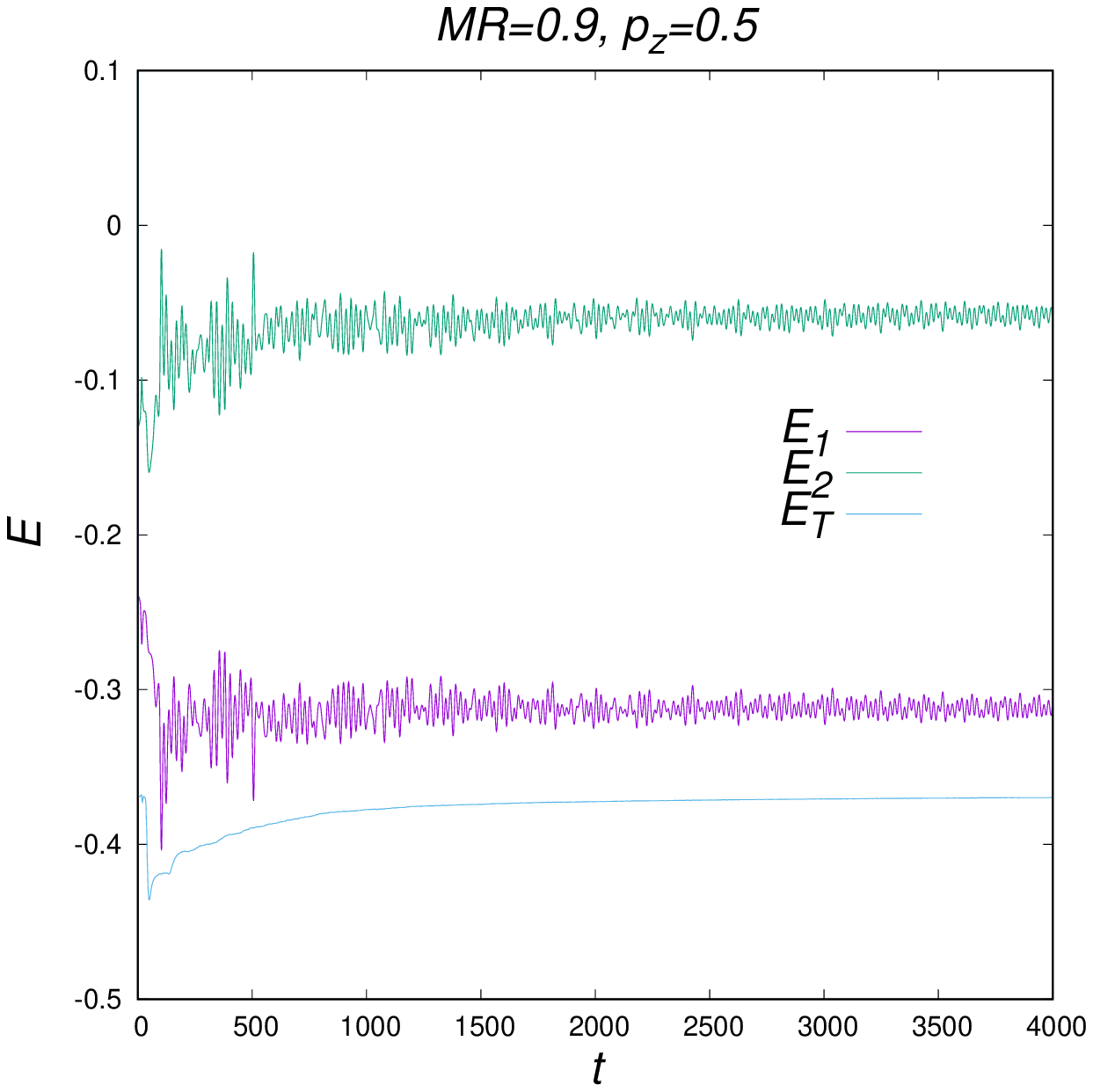} 
\includegraphics[width=2.5cm]{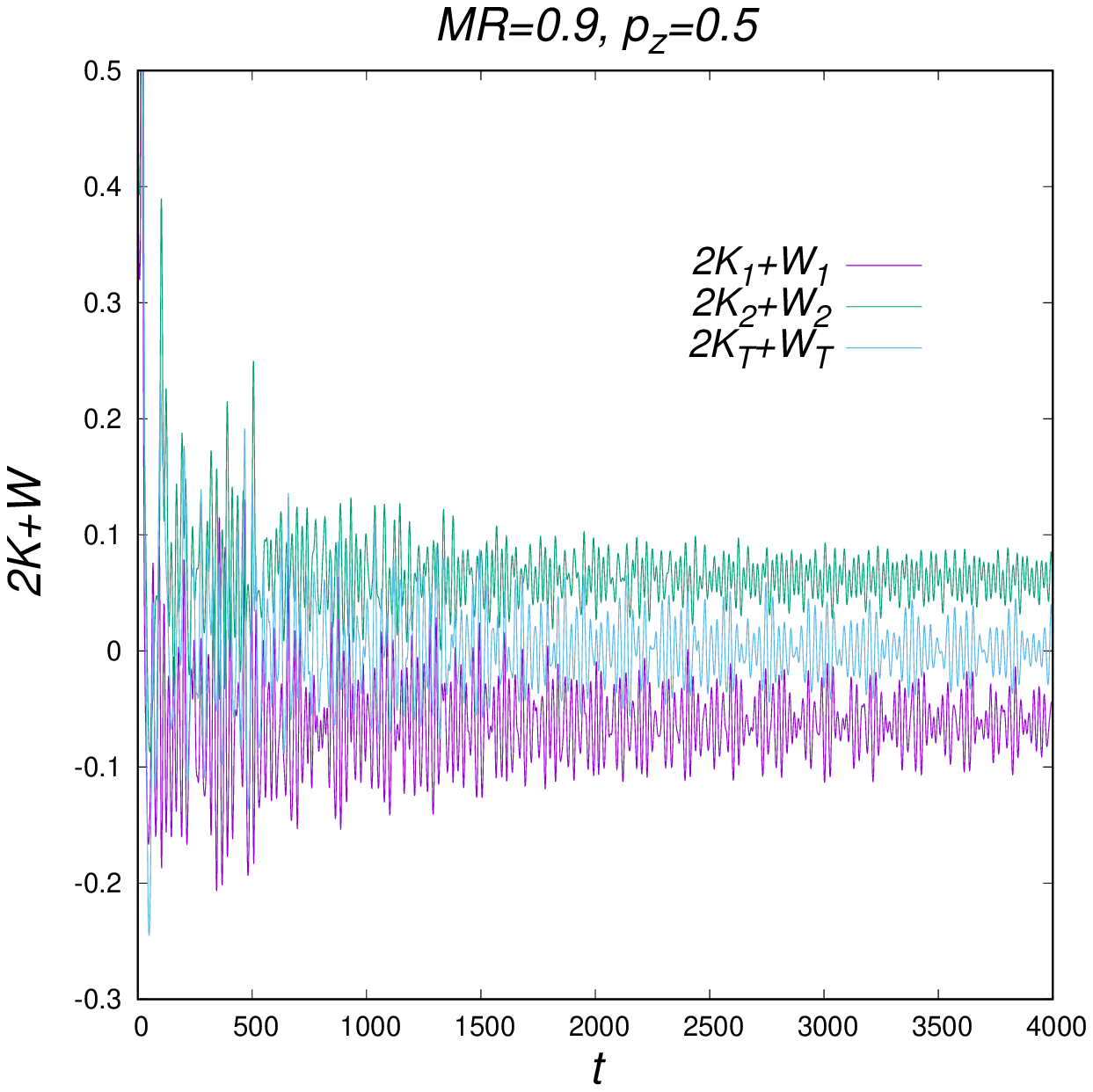} 
\includegraphics[width=2.5cm]{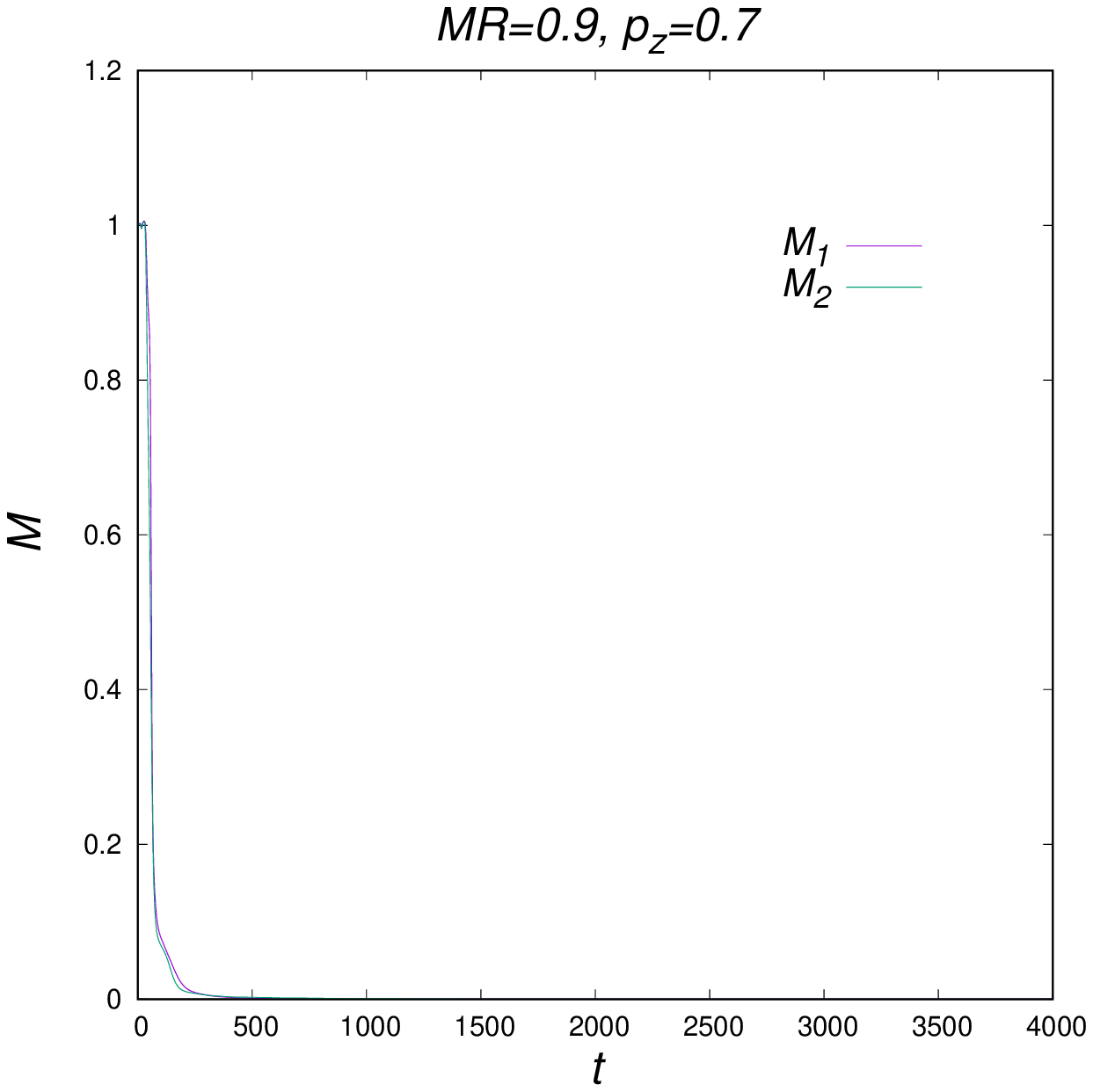} 
\includegraphics[width=2.5cm]{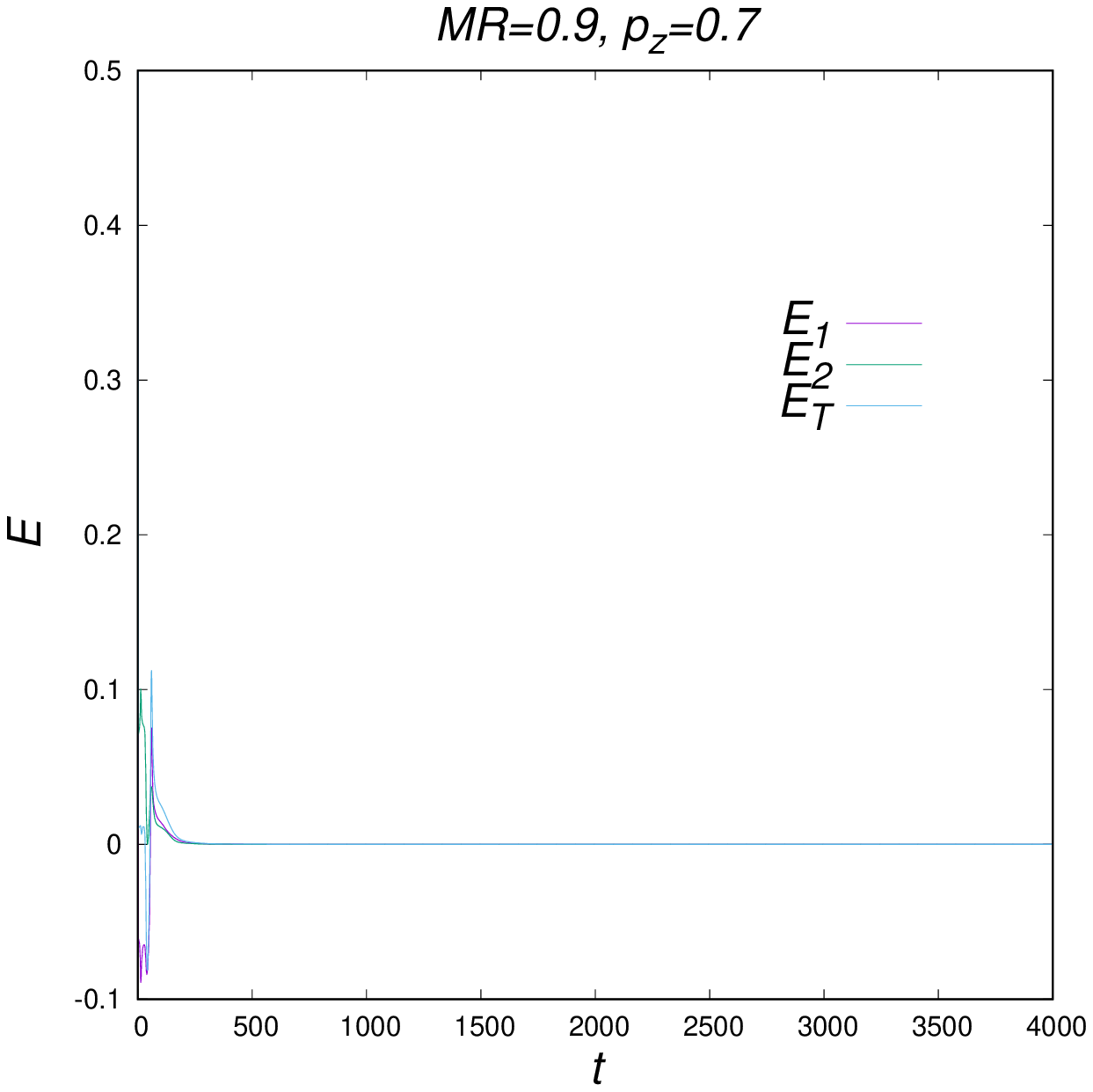} 
\includegraphics[width=2.5cm]{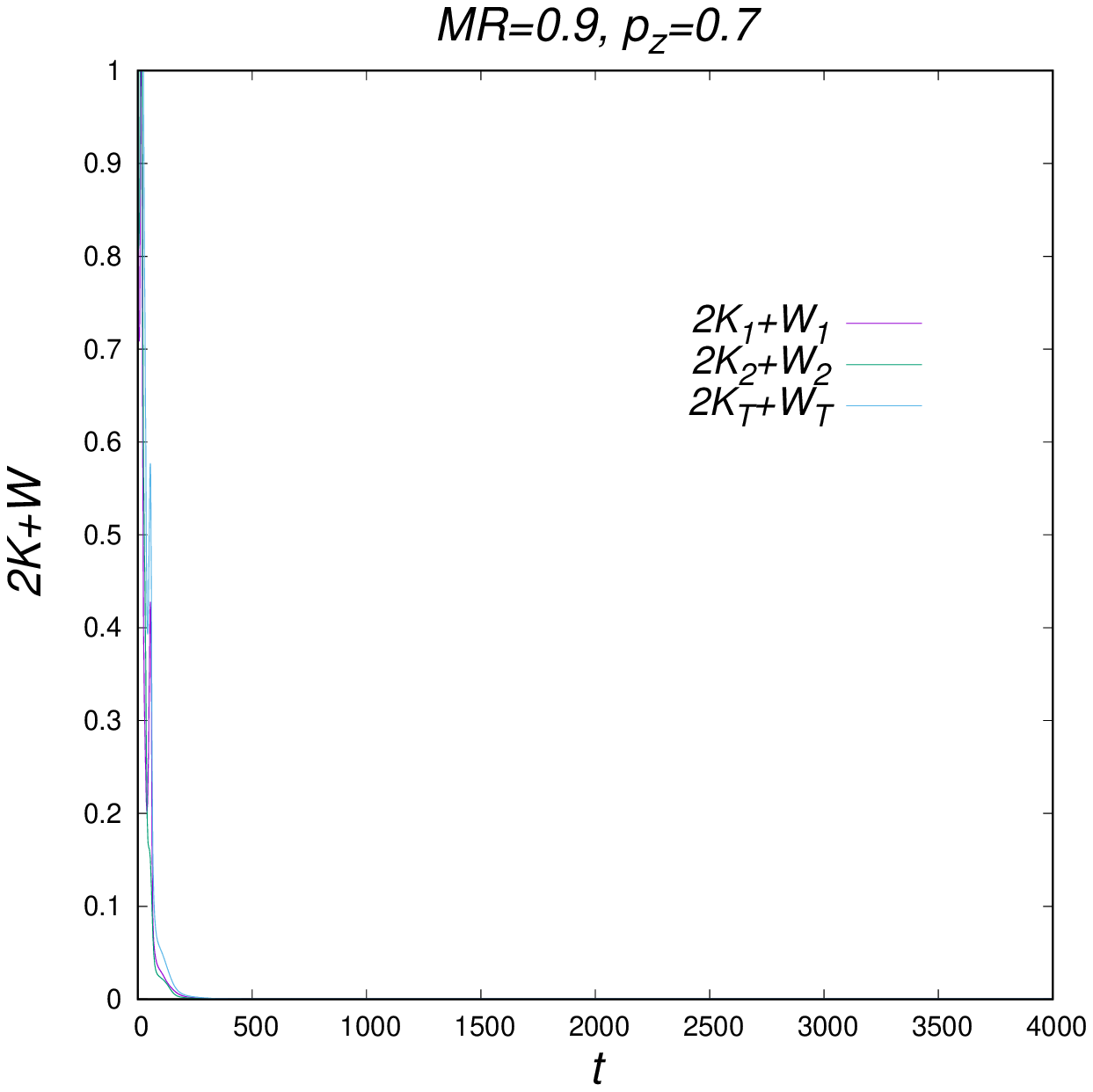} 
\caption{\label{fig:mr0_9_pz0_2_scalars} (Left) Mass of each state as function of time, normalized with the initial value. (Middle) Energy of each state and $E_T$. (Right) The quantity $2K_i+W_i$ for each state and the two together. These scalars corresponds to the case $MR=0.9$ and $p_z=0.2,0.5,0.7$.}
\end{figure}

During the evolution we measure the mass of each state $M_1=\int \rho_1 d^3 x$ and $M_2=\int \rho_2 d^3 x$ that we use to monitor the evolution of the configuration. In particular we simulate the evolution during a time window that allows the masses of the states to stabilize as in the left column of Fig. \ref{fig:mr0_9_pz0_2_scalars}. 

The kinetic energy $K_i=\int \Psi^*_i (-\frac{1}{2}\nabla^2)\Psi d^3x$ and the gravitational energy $W_i=\int \Psi^* V \Psi d^3 x$ of each state $i=1,2$ are also useful quantities to characterize the evolution of the system. 
Each state has total energy $E_i=K_i+W_i$ and  the total energy of the system is $E_T=K_T+W_T$, where $K_T=K_1+K_2$ and $W_T=W_1+W_2$. In the middle column of Fig. \ref{fig:mr0_9_pz0_2_scalars} we show the energy of each state that oscillates with high frequency and the total energy that is smooth and approaches a constant value. In the right column we show the quantity $2K_i+W_i$ for each state separately and also for the whole system. The result is that when considered together, $2K_T + W_T$ oscillates with decreasing amplitude around zero, indicating that the system as a whole, approaches a virialized state.

The final mass ratio is different from $MR$; it depends on the initial head-on momentum $p_{z}$ and on the nonlinearities of the evolution that expel different amounts of mass from the domain during the merger as seen in Fig.  \ref{fig:mr0_9_pz0_2_scalars}. For example, for the case with $MR=0.9$ and $p_z=0.2$ the final mass ratio approaches $MR_{final}\sim 0.693$, whereas for $MR=0.9$ and $p_z=0.3$ the final mass ratio is $MR_{final}\sim 0.715$. Also in this Figure we show the case with $p_{z}=0.7$ that has a slightly positive total energy at initial time and the result is the escape of all the mass of the two configurations through the boundary.

\begin{figure}[htp]
\includegraphics[width=2.5cm]{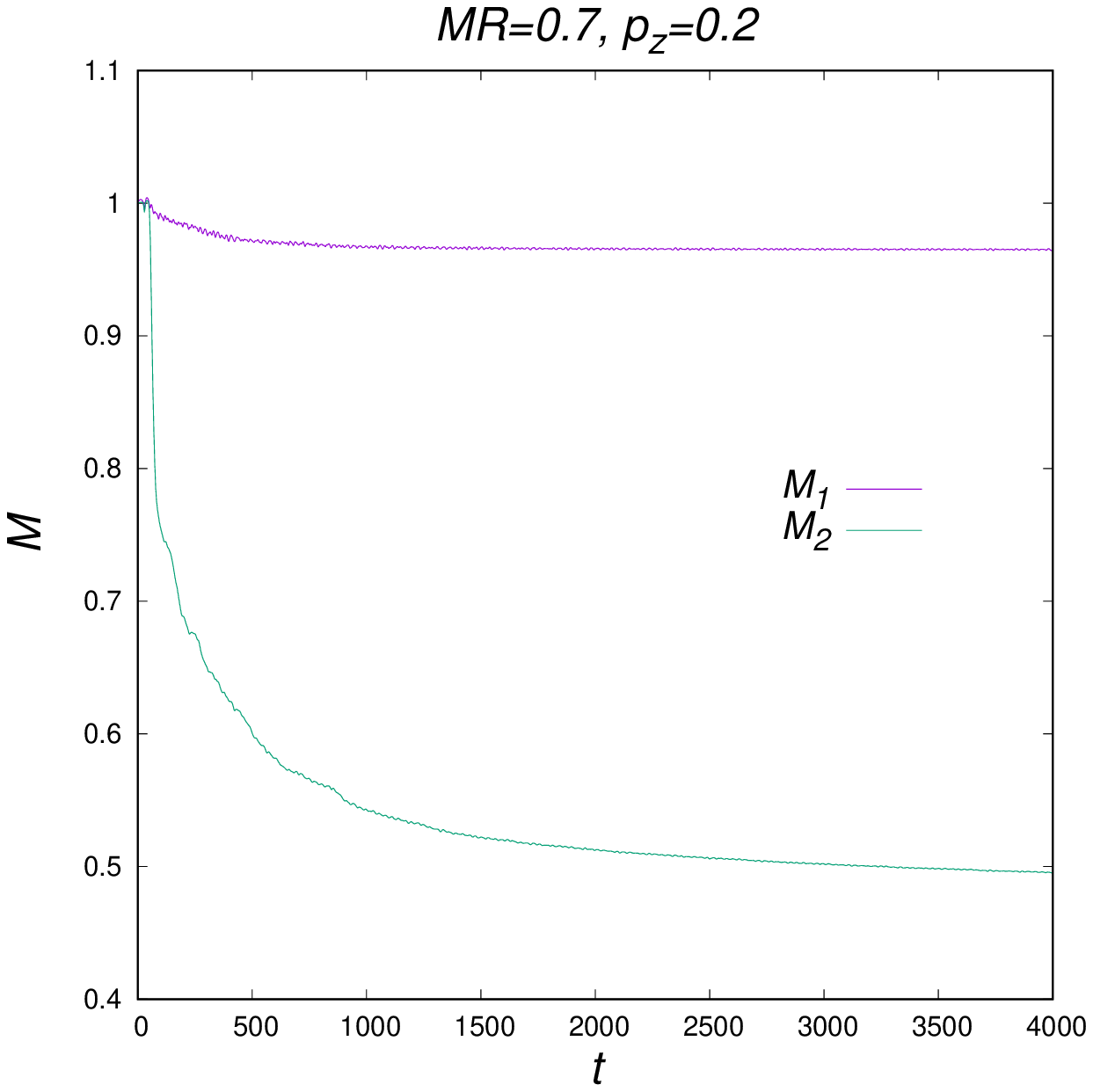} 
\includegraphics[width=2.5cm]{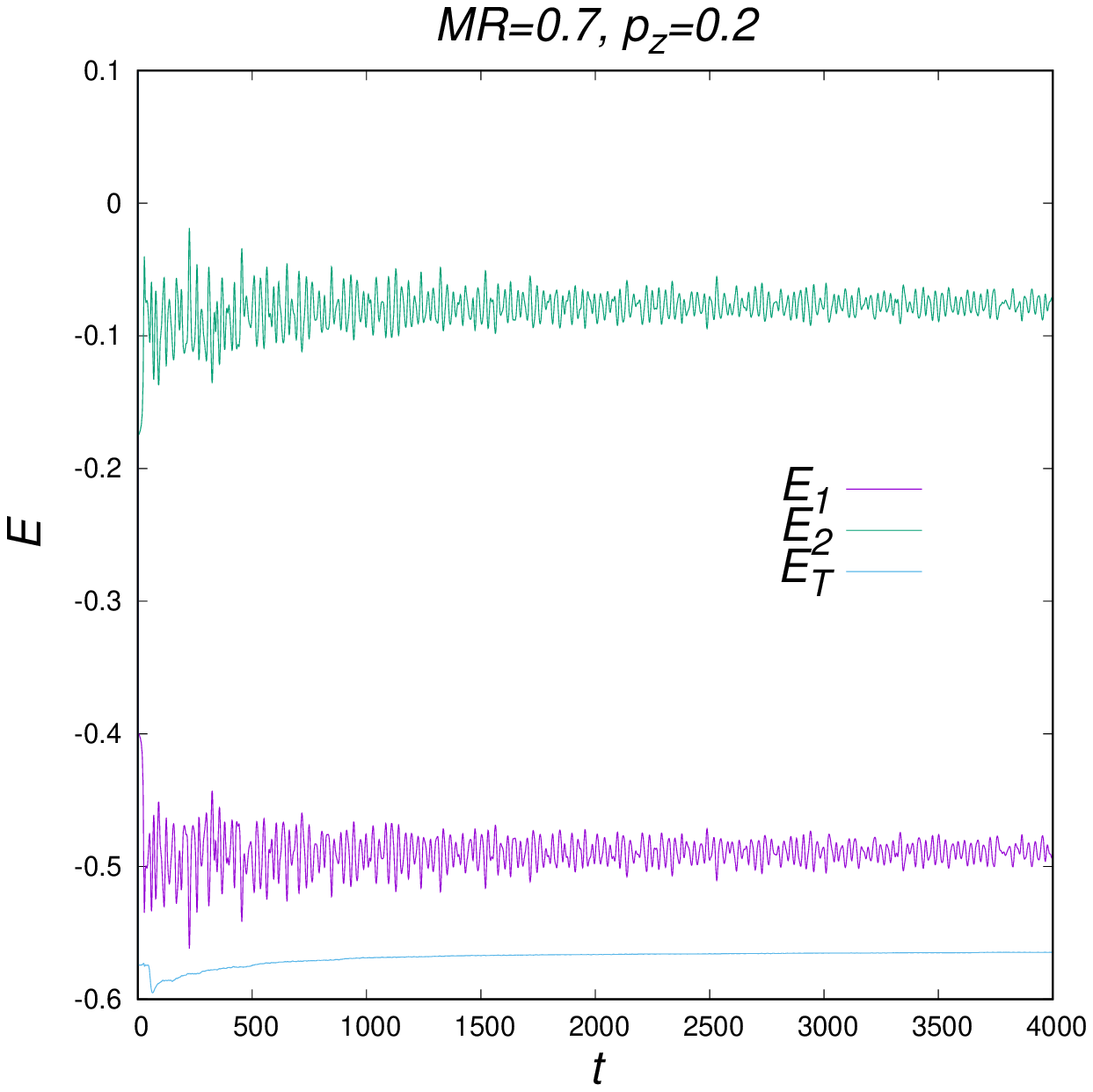} 
\includegraphics[width=2.5cm]{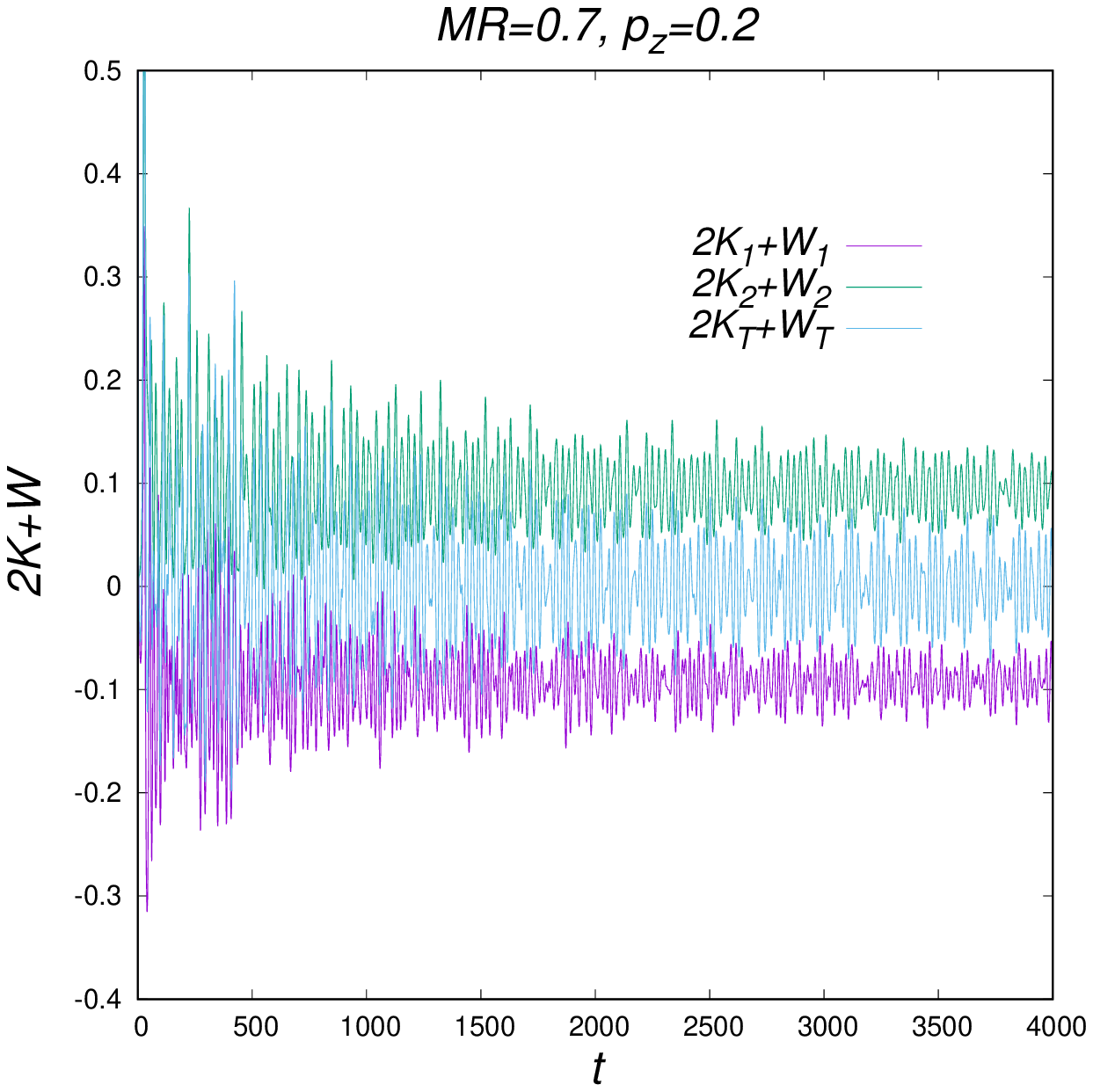} 
\includegraphics[width=2.5cm]{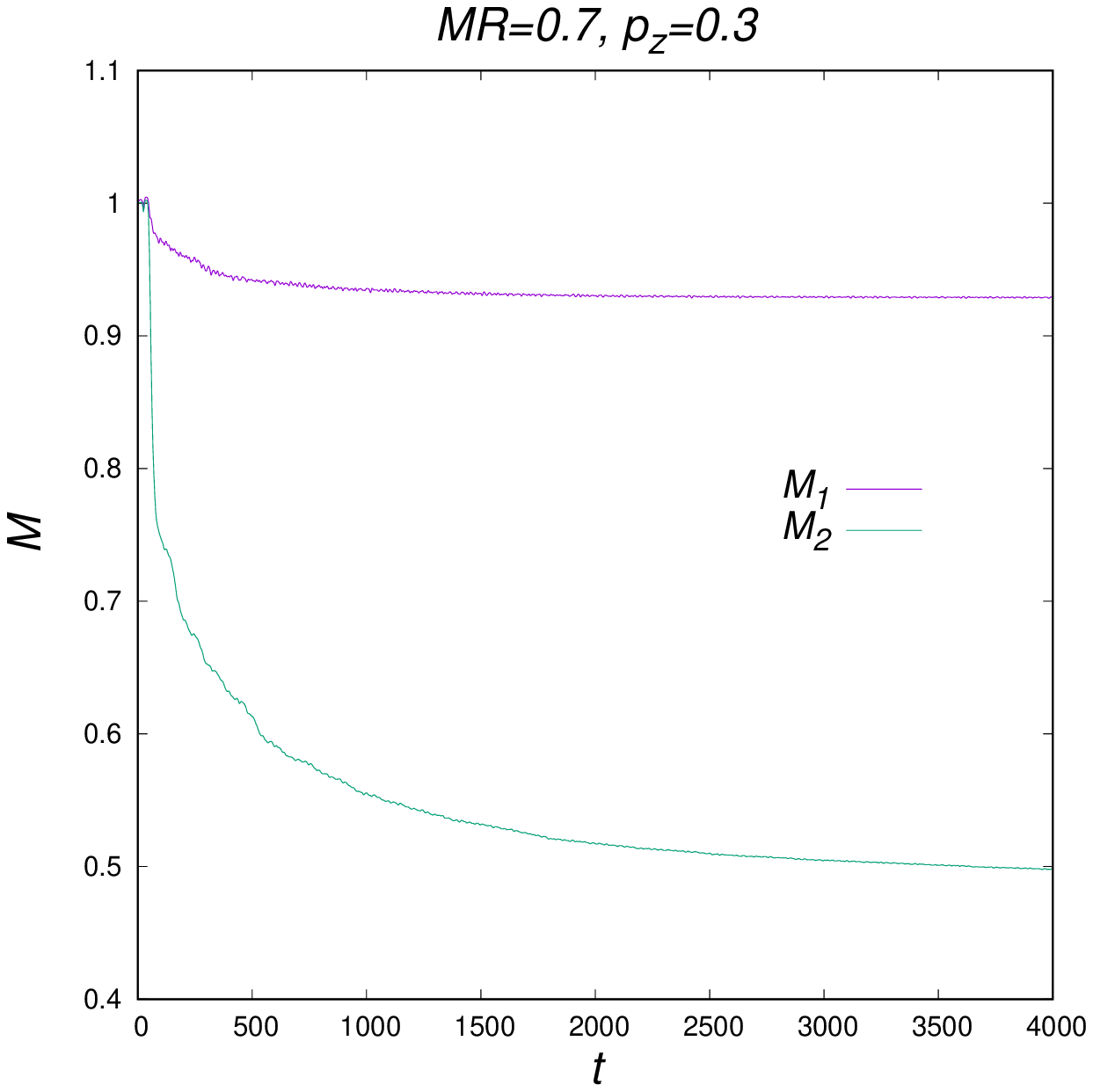} 
\includegraphics[width=2.5cm]{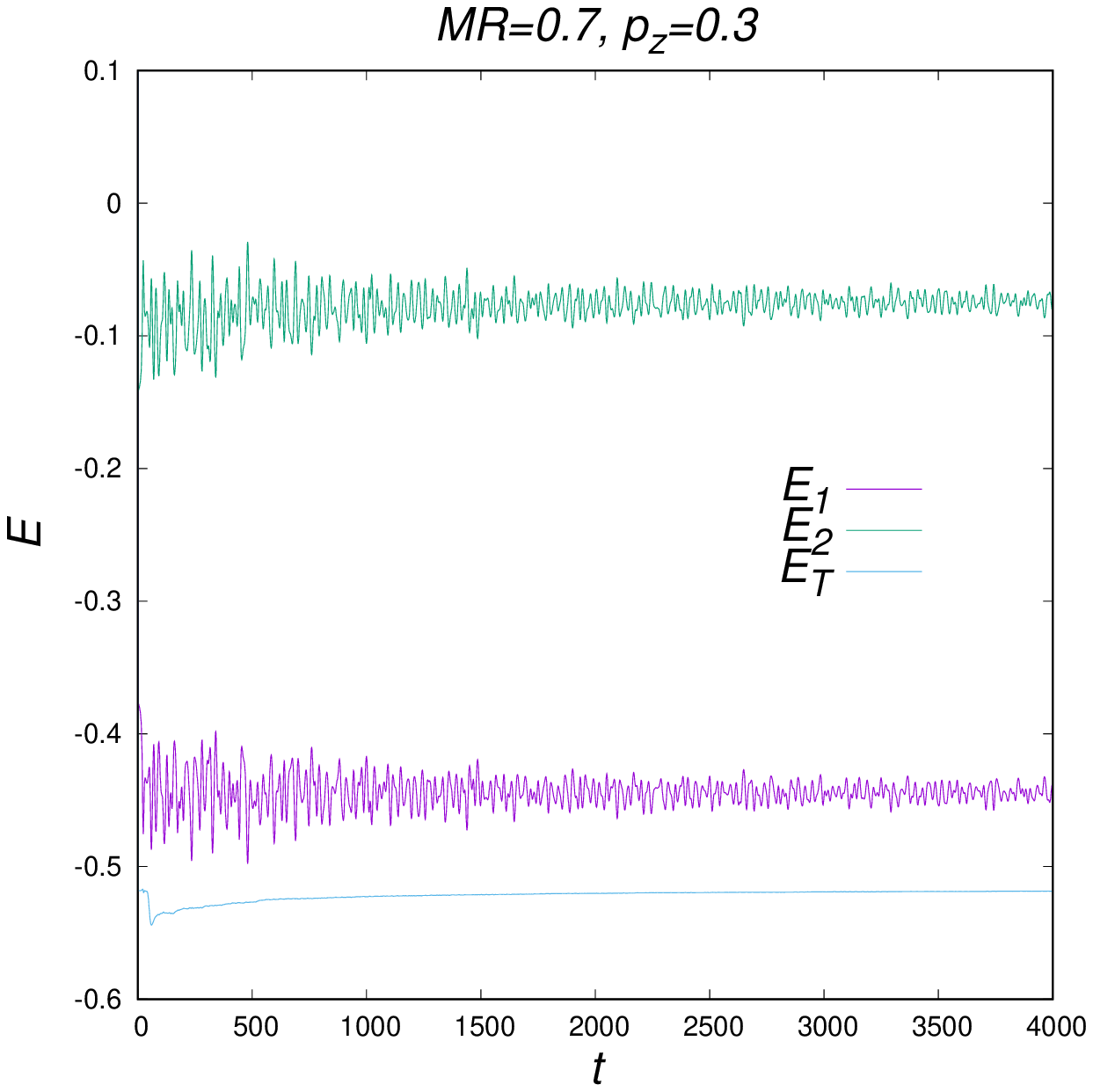} 
\includegraphics[width=2.5cm]{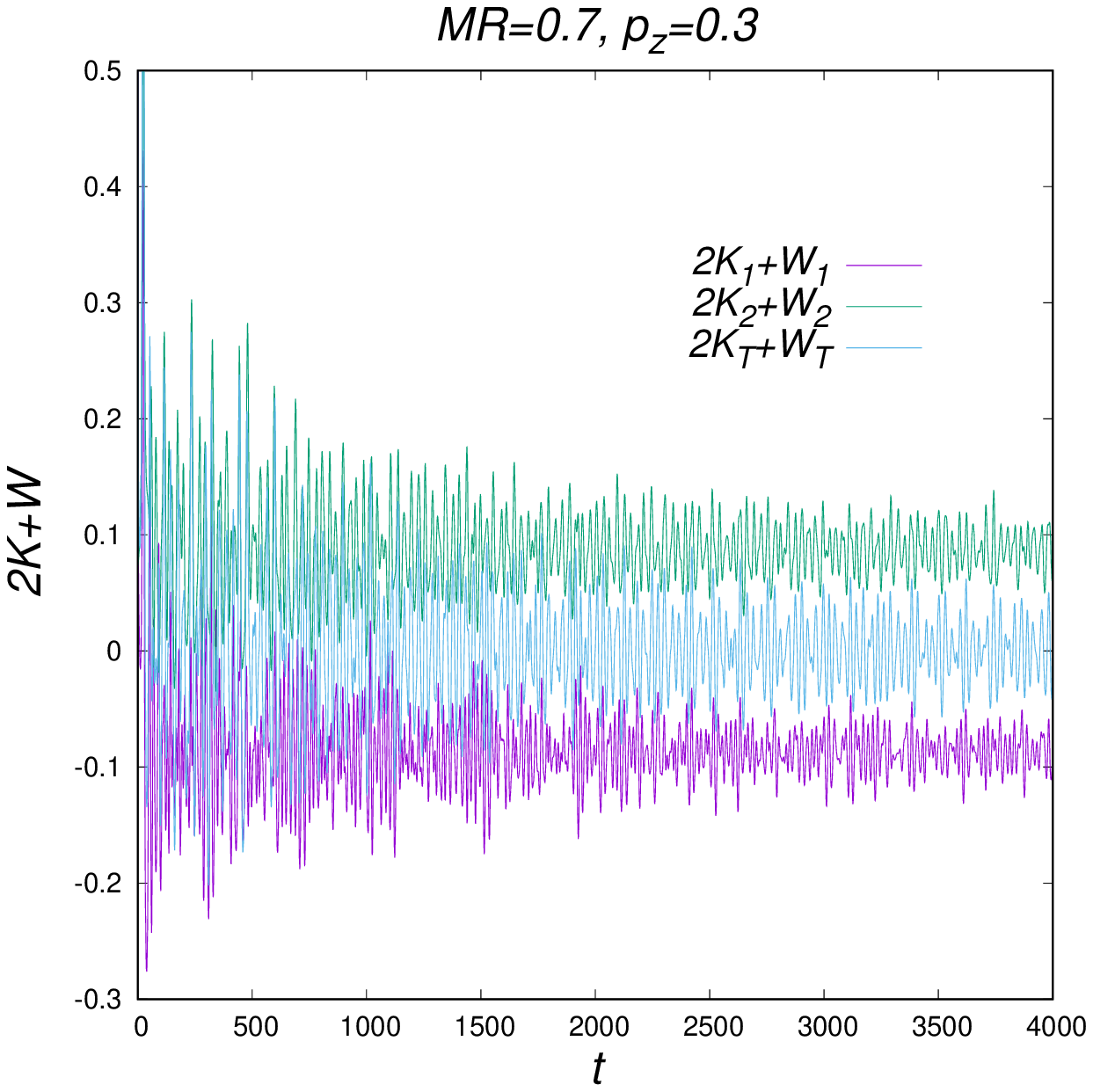} 
\caption{\label{fig:mr0_7_scalars} (Left) Mass of each state as function of time, normalized with the initial value. (Middle) Energy of each state and $E_T$. (Right) The quantity $2K+W$ for each state and the two together. These scalars correspond to the case $MR=0.7$ and $p_z=0.2,0.3$.}
\end{figure}

\begin{figure}[htp]
\includegraphics[width=2.5cm]{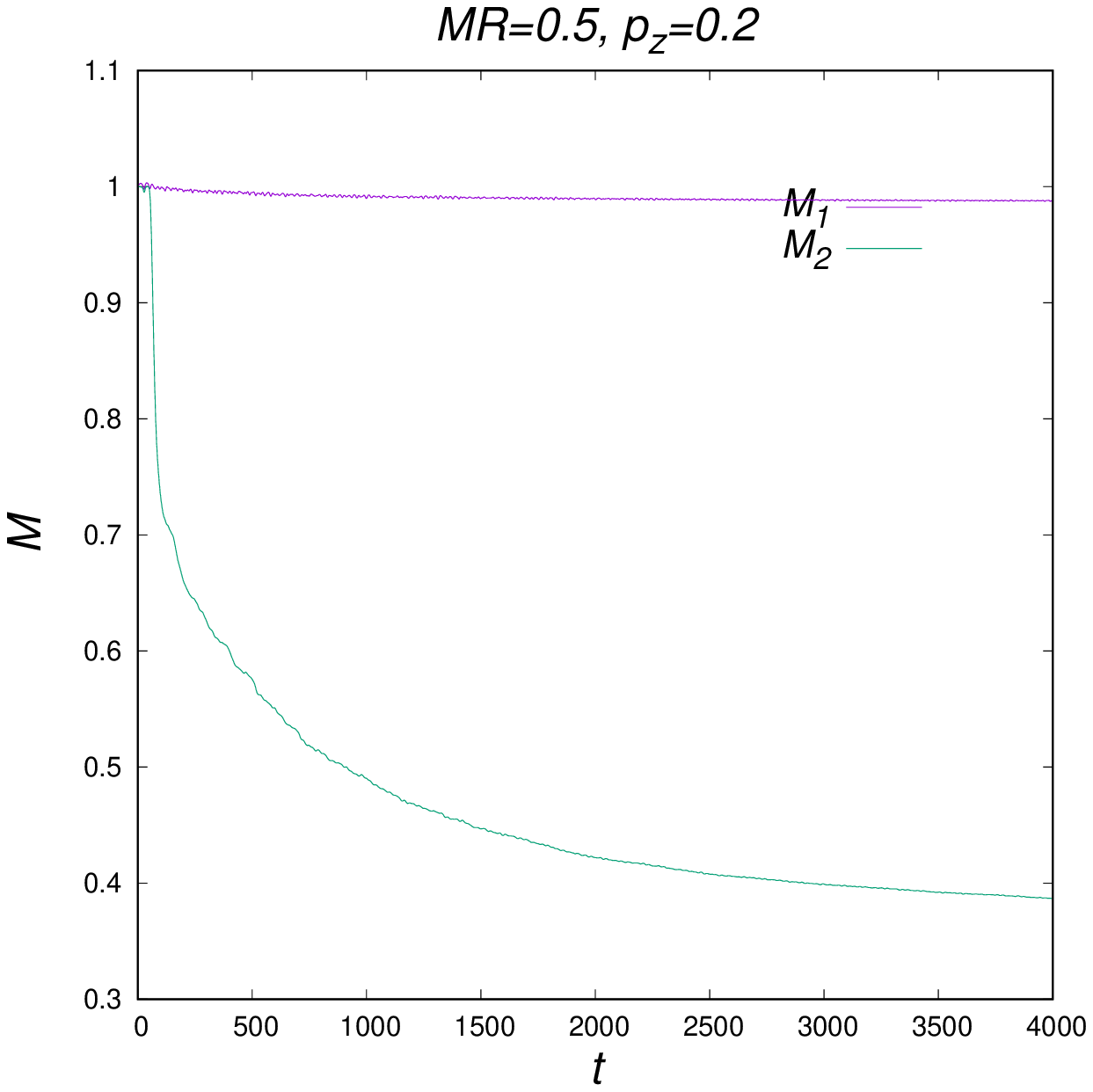} 
\includegraphics[width=2.5cm]{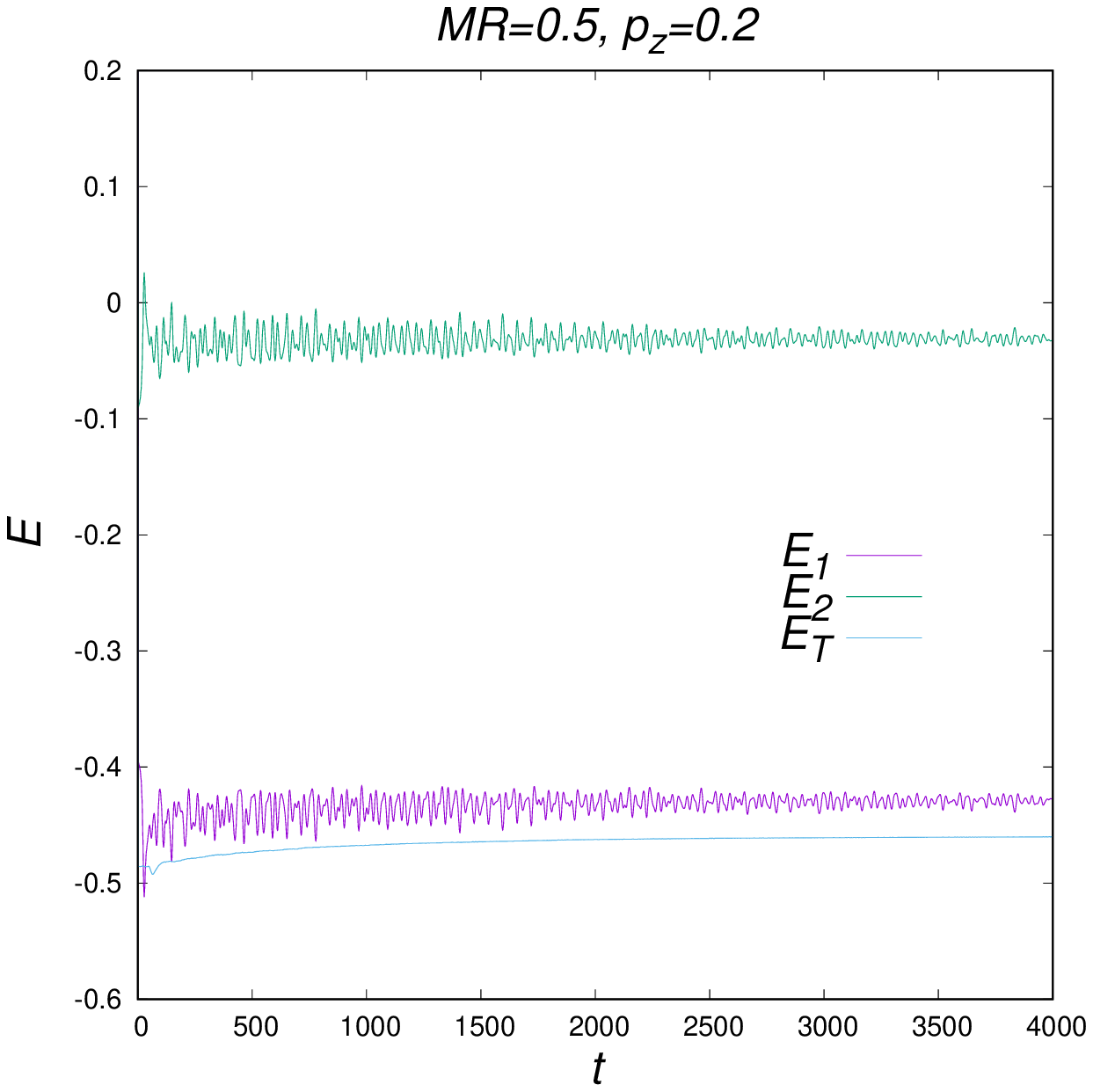} 
\includegraphics[width=2.5cm]{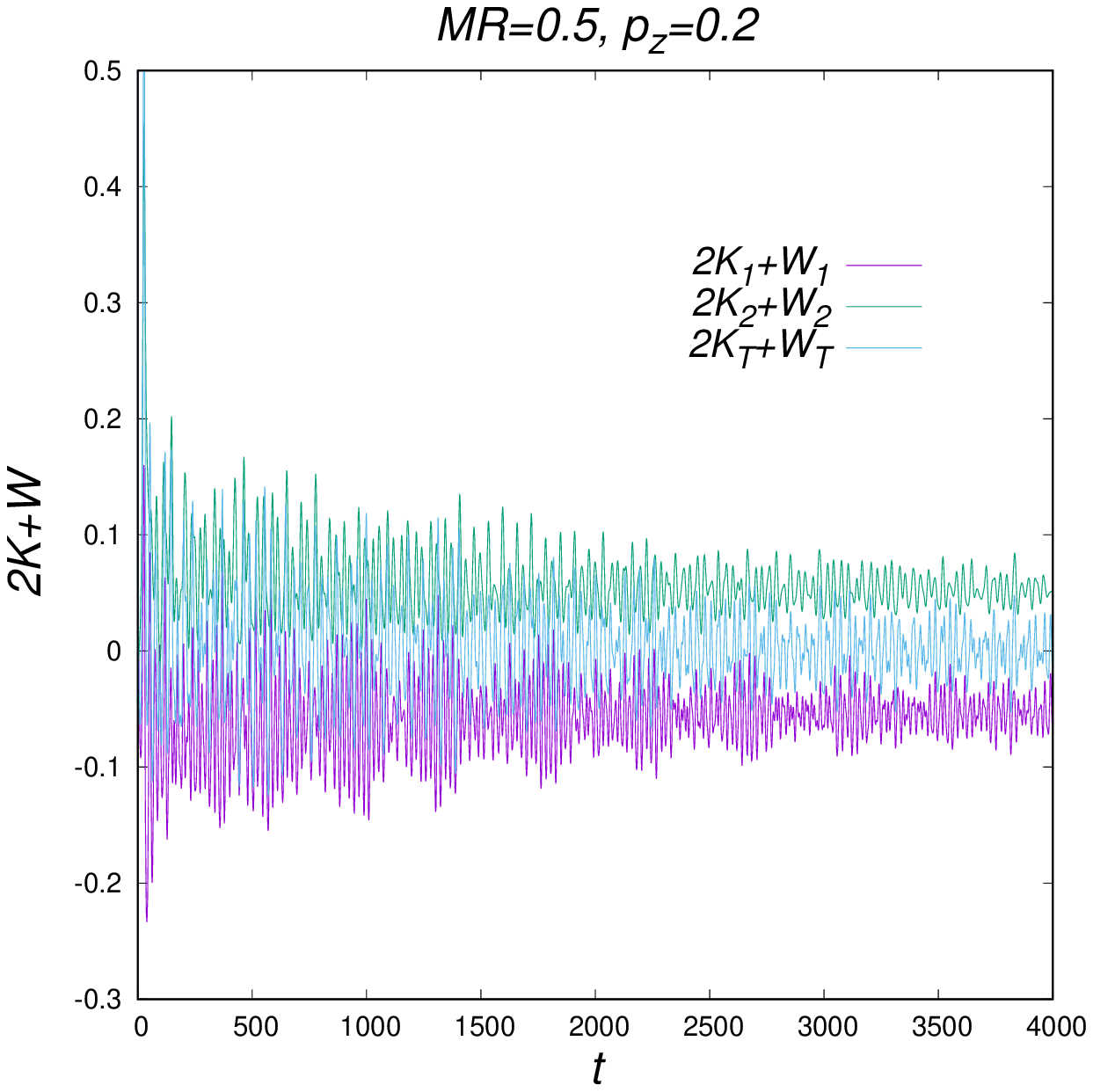} 
\includegraphics[width=2.5cm]{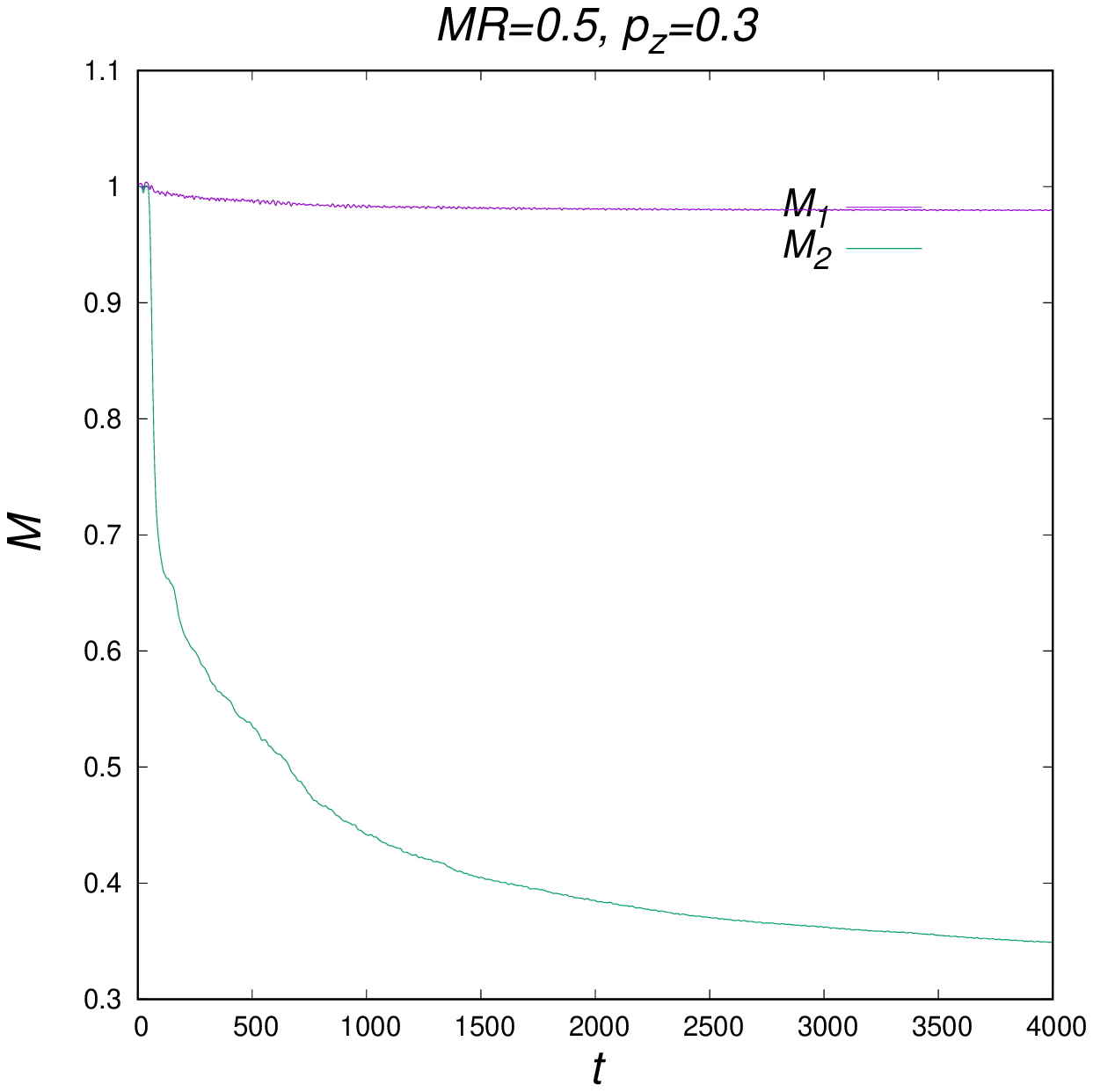} 
\includegraphics[width=2.5cm]{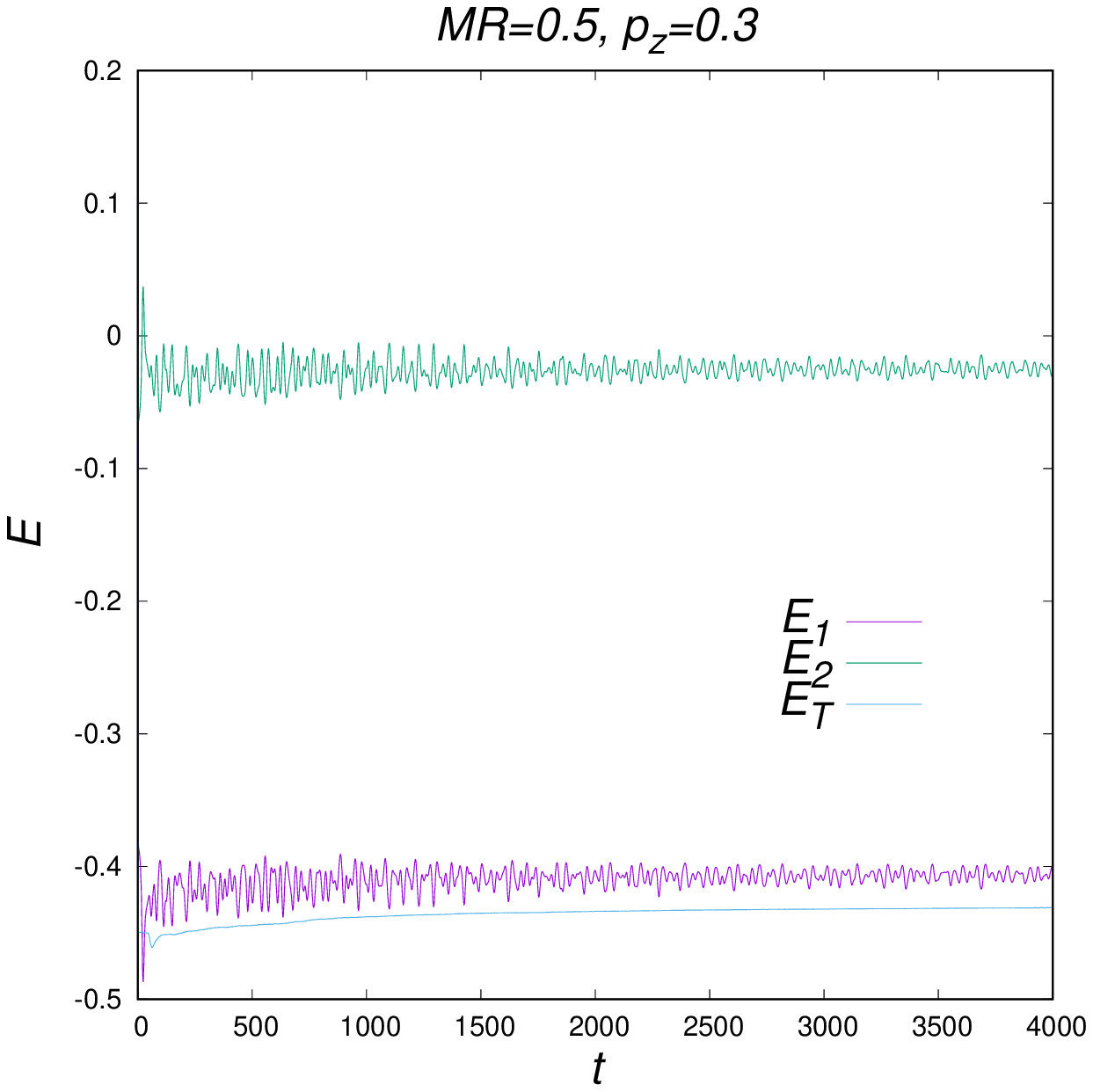} 
\includegraphics[width=2.5cm]{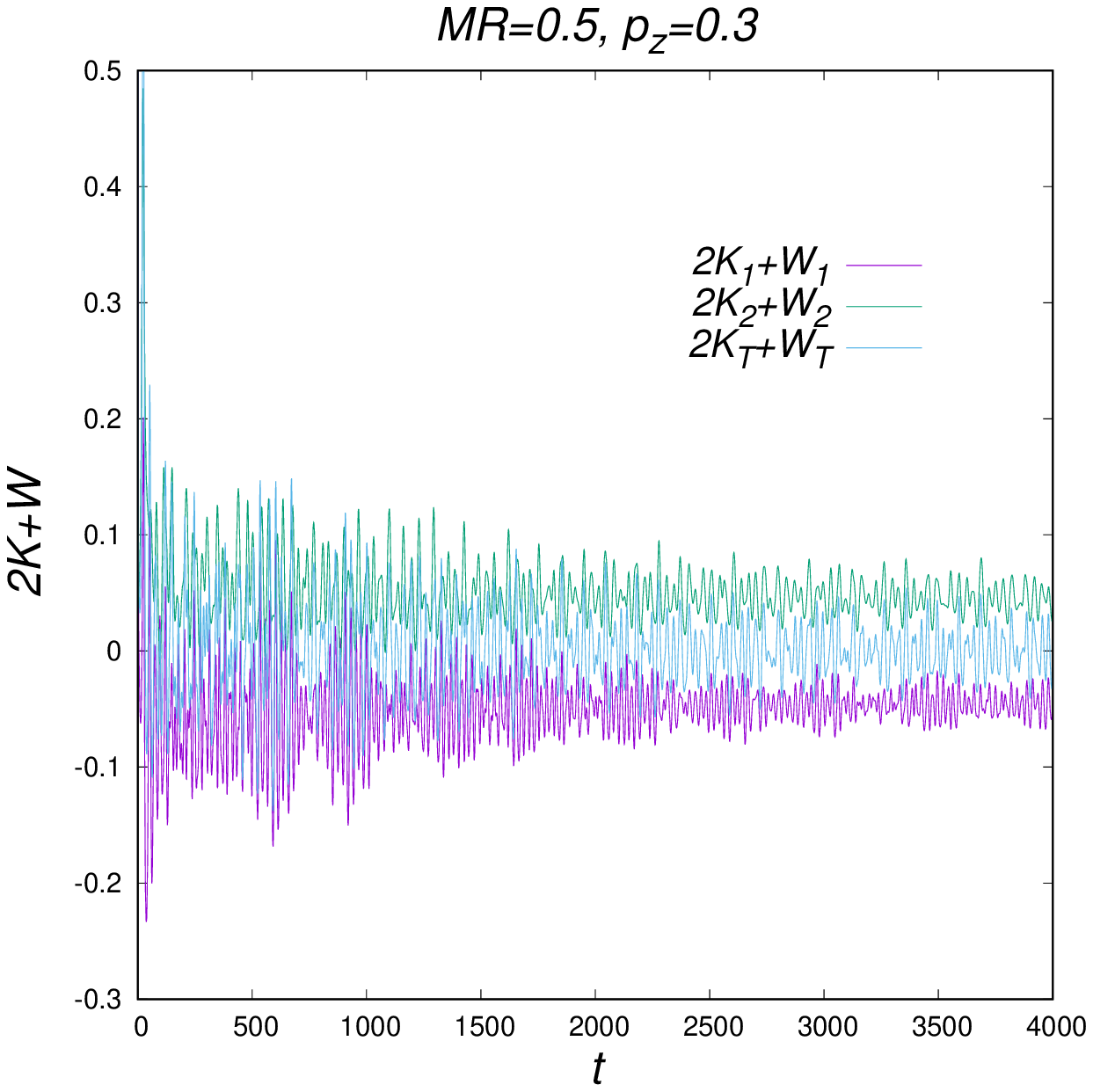} 
\caption{\label{fig:mr0_5_scalars} (Left) Mass of each state as function of time, normalized with the initial value. (Middle) Energy of each state and $E_T$. (Right) The quantity $2K+W$ for each state and the two together. These scalars correspond to the case $MR=0.5$ and $p_z=0.2,0.3$.}
\end{figure}

A summary of the evolution is that the result of the merger shows the morphology of the mixed $(1,0,0)+(2,1,0)$ configuration, it does not seem to stop oscillating around the coordinate origin, and the distribution of the less massive state does not become stationary as seen in the animations. On the other hand, despite the highly dynamical behavior, it seems the system approaches a virialized state according to the results in Fig. \ref{fig:mr0_9_pz0_2_scalars}. The behavior is generic in our parameter space, which is illustrated in Fig. \ref{fig:mr0_7_scalars} and \ref{fig:mr0_5_scalars} for the cases with $MR=0.7$ and 0.5.

\begin{figure}[htp]
\includegraphics[width=2.5cm]{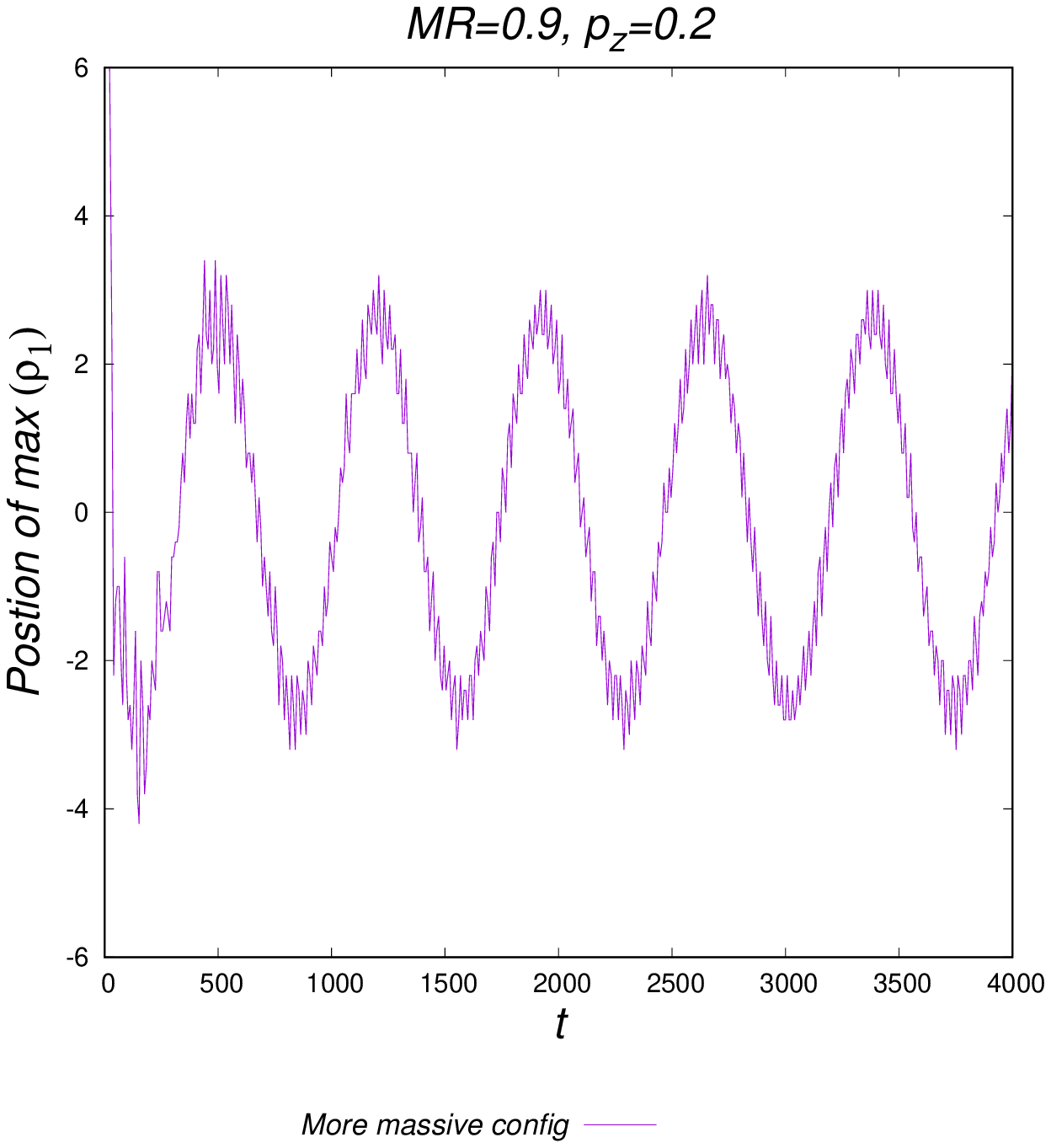} 
\includegraphics[width=2.5cm]{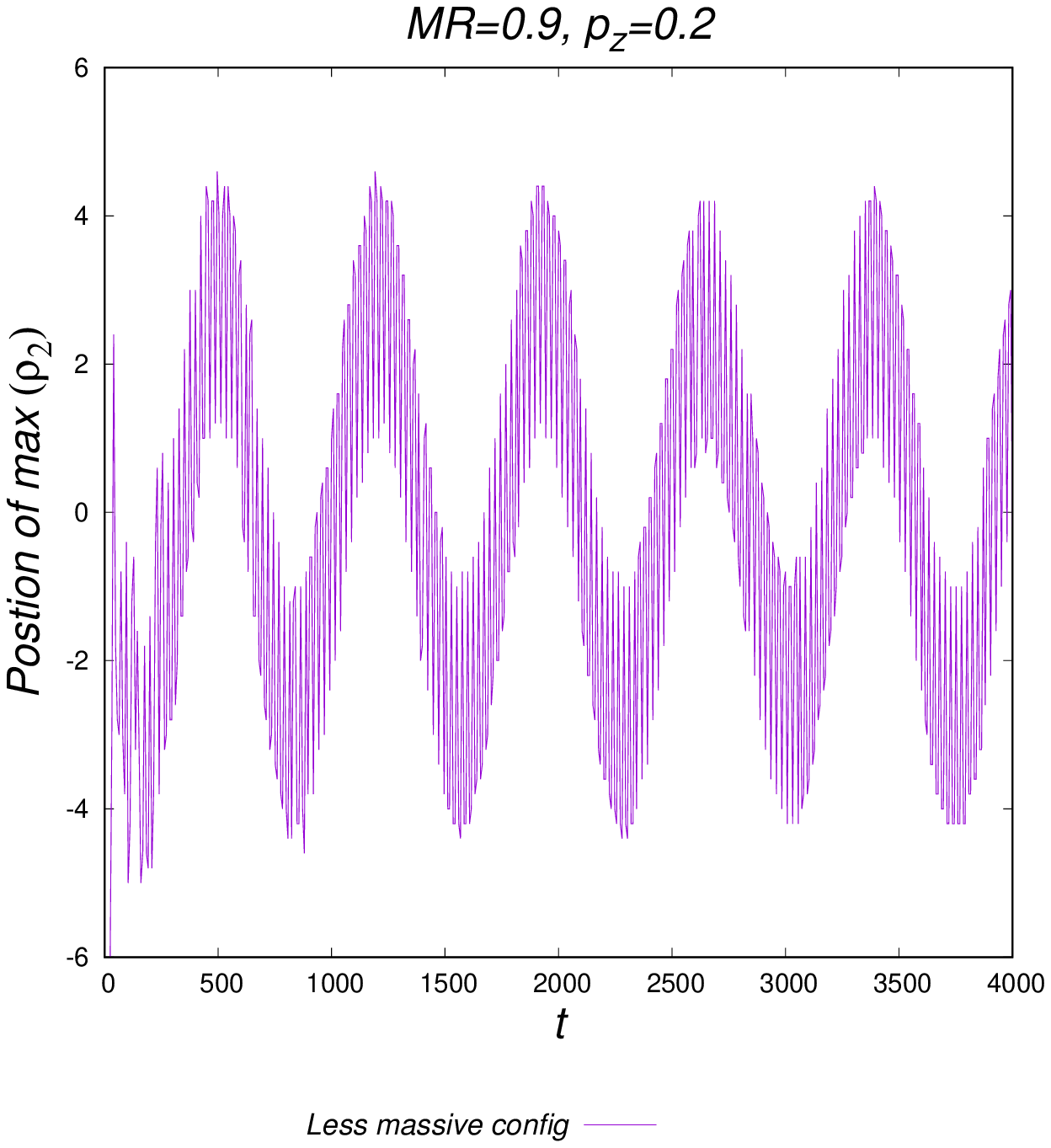} 
\includegraphics[width=2.5cm]{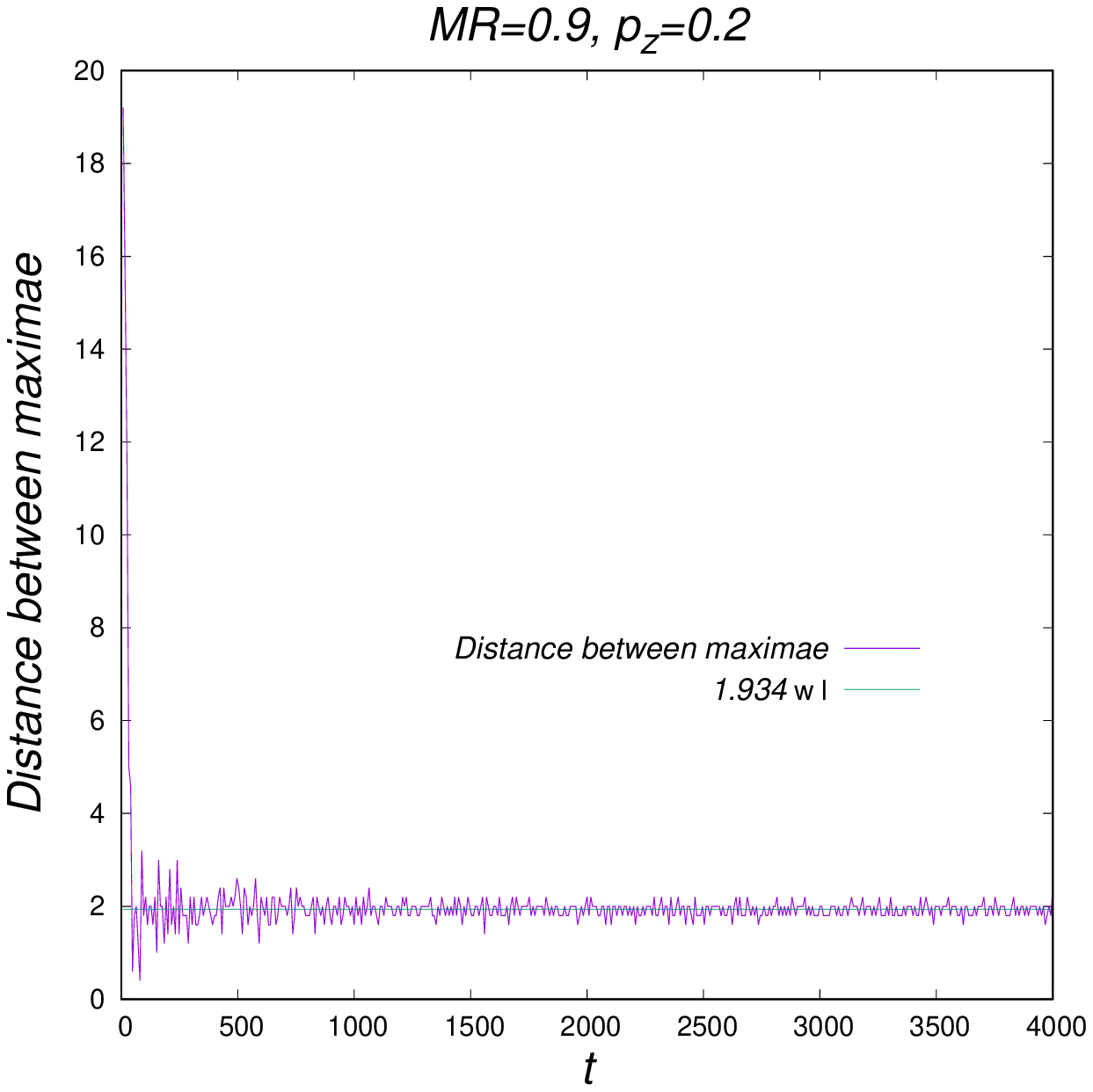} 
\includegraphics[width=2.5cm]{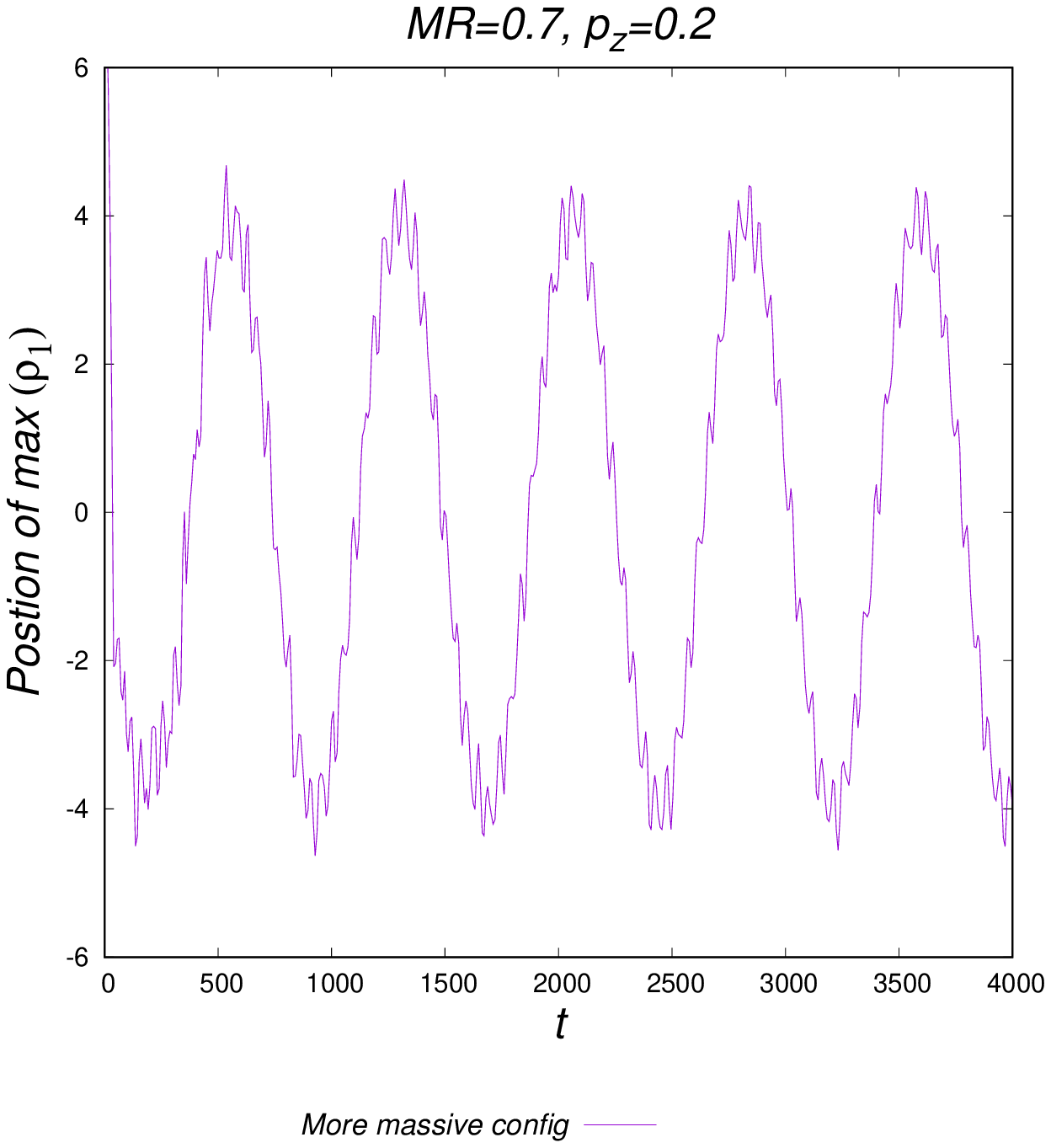} 
\includegraphics[width=2.5cm]{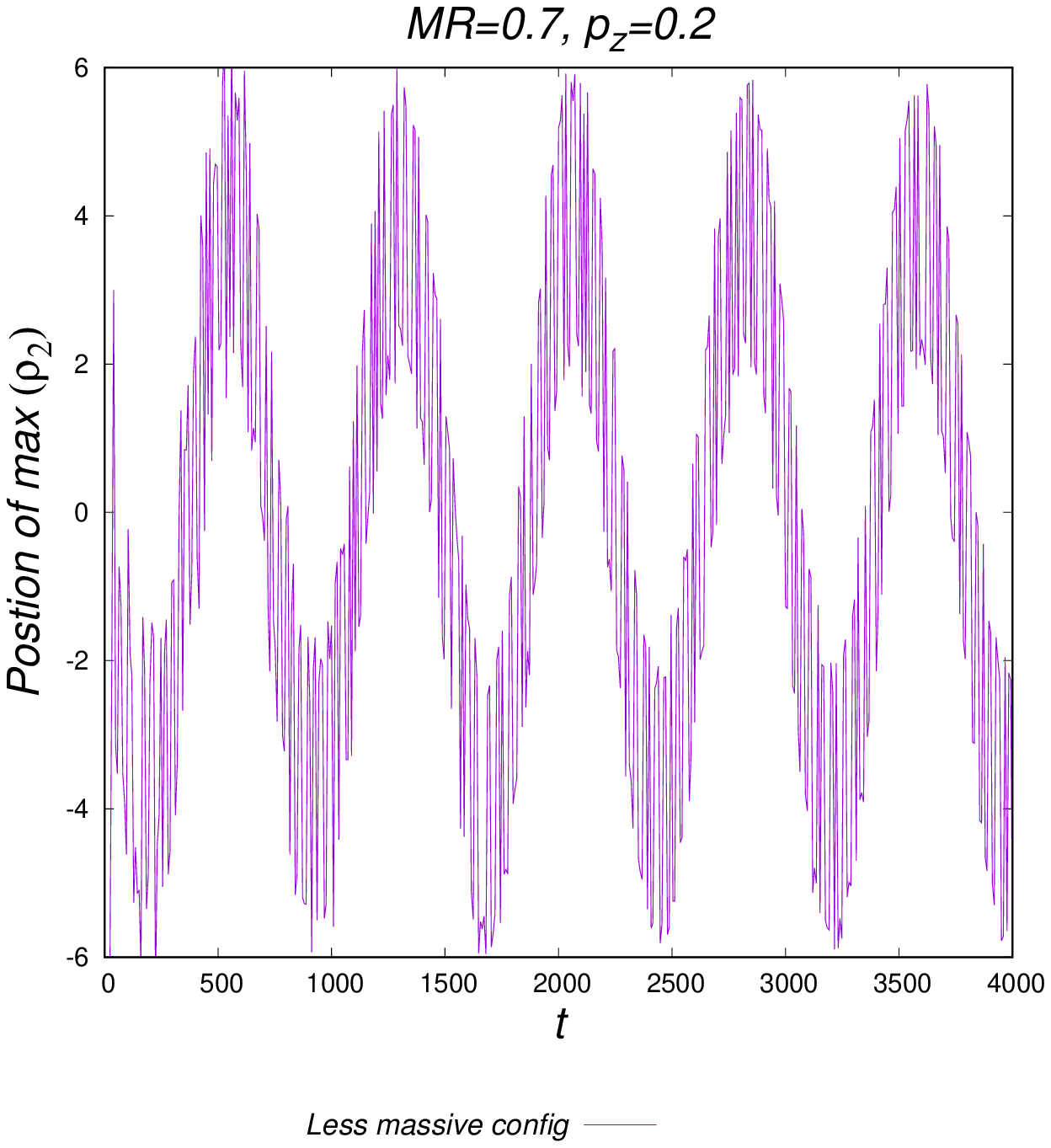} 
\includegraphics[width=2.5cm]{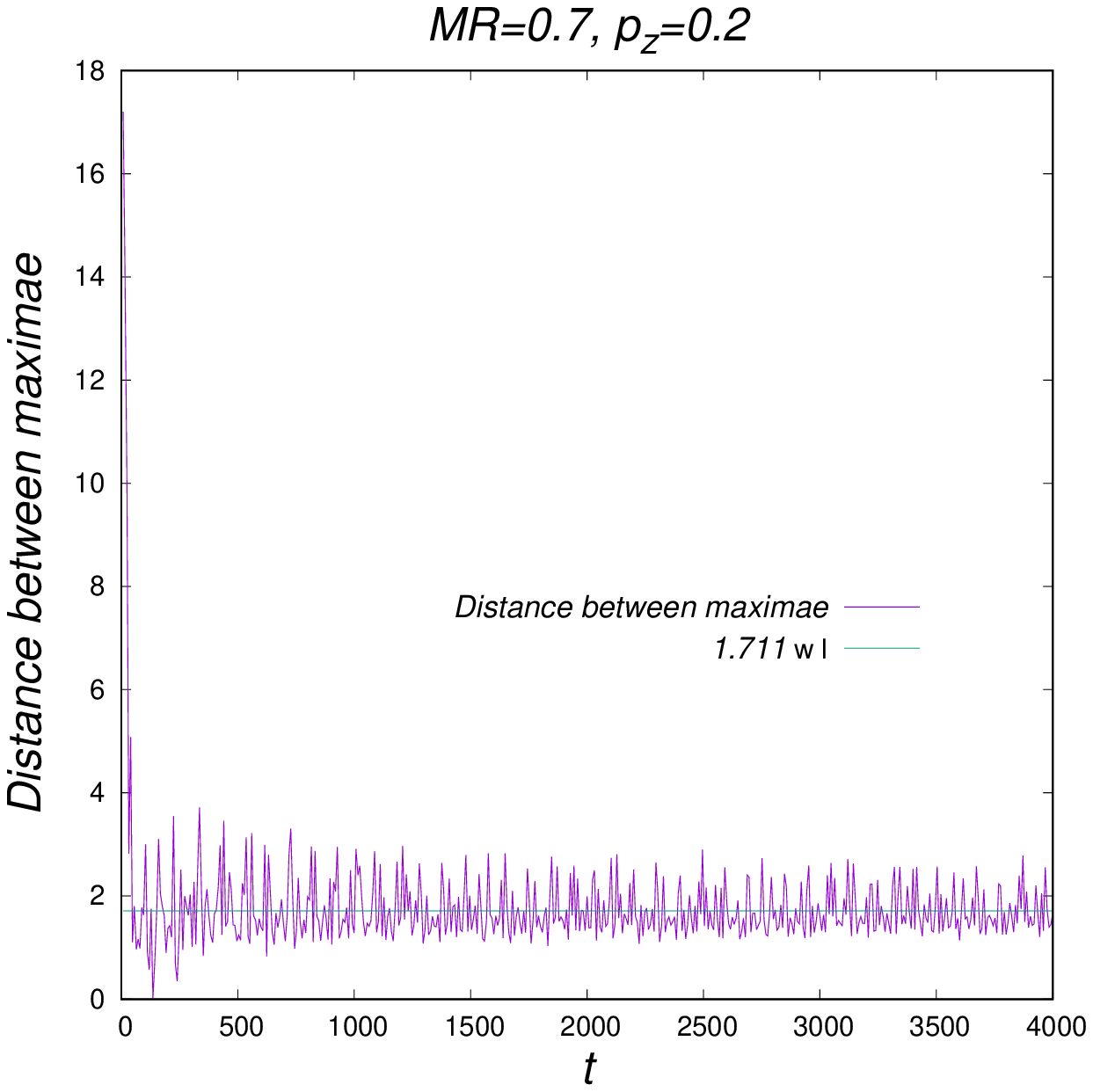} 
\includegraphics[width=2.5cm]{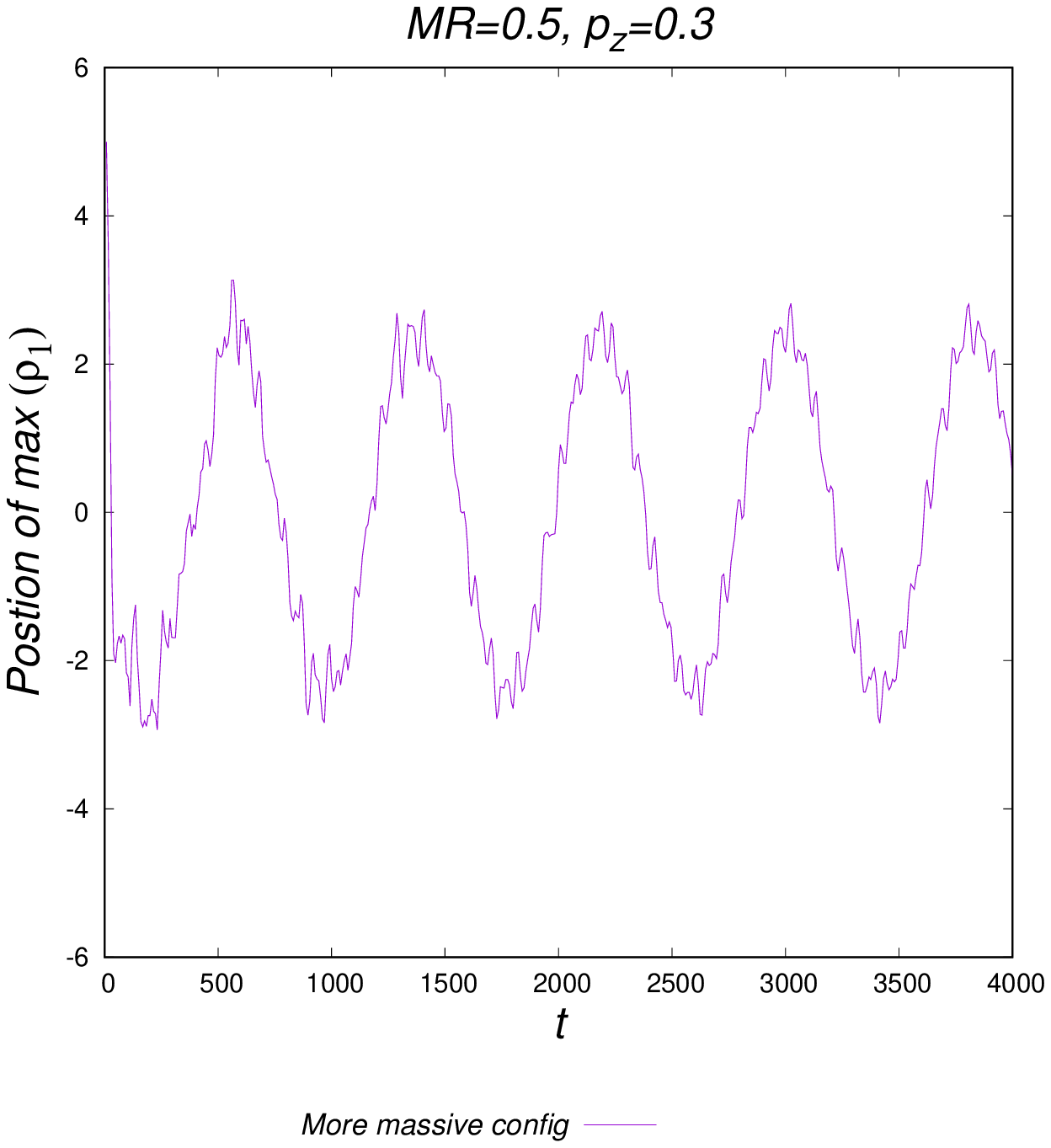} 
\includegraphics[width=2.5cm]{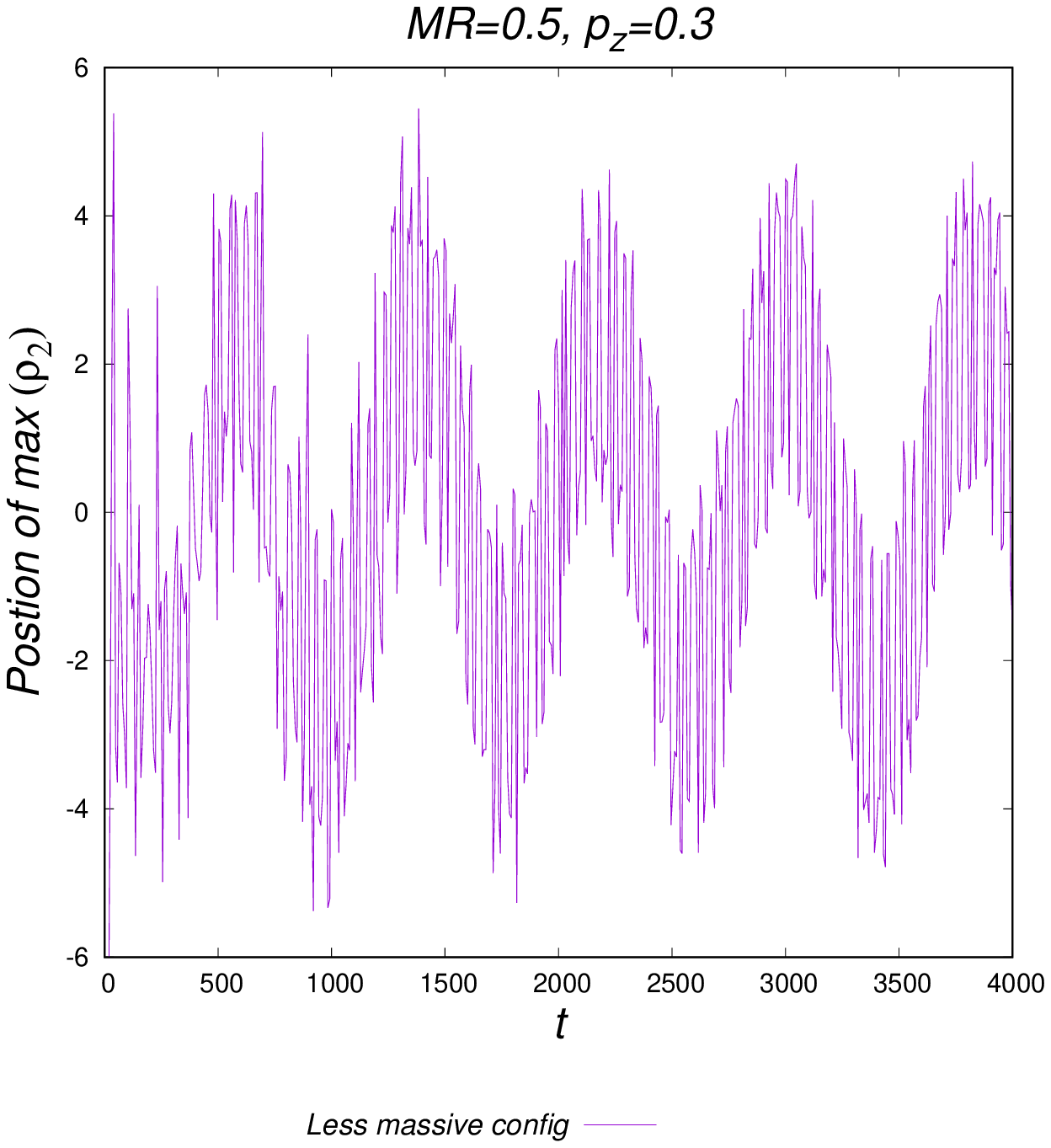} 
\includegraphics[width=2.5cm]{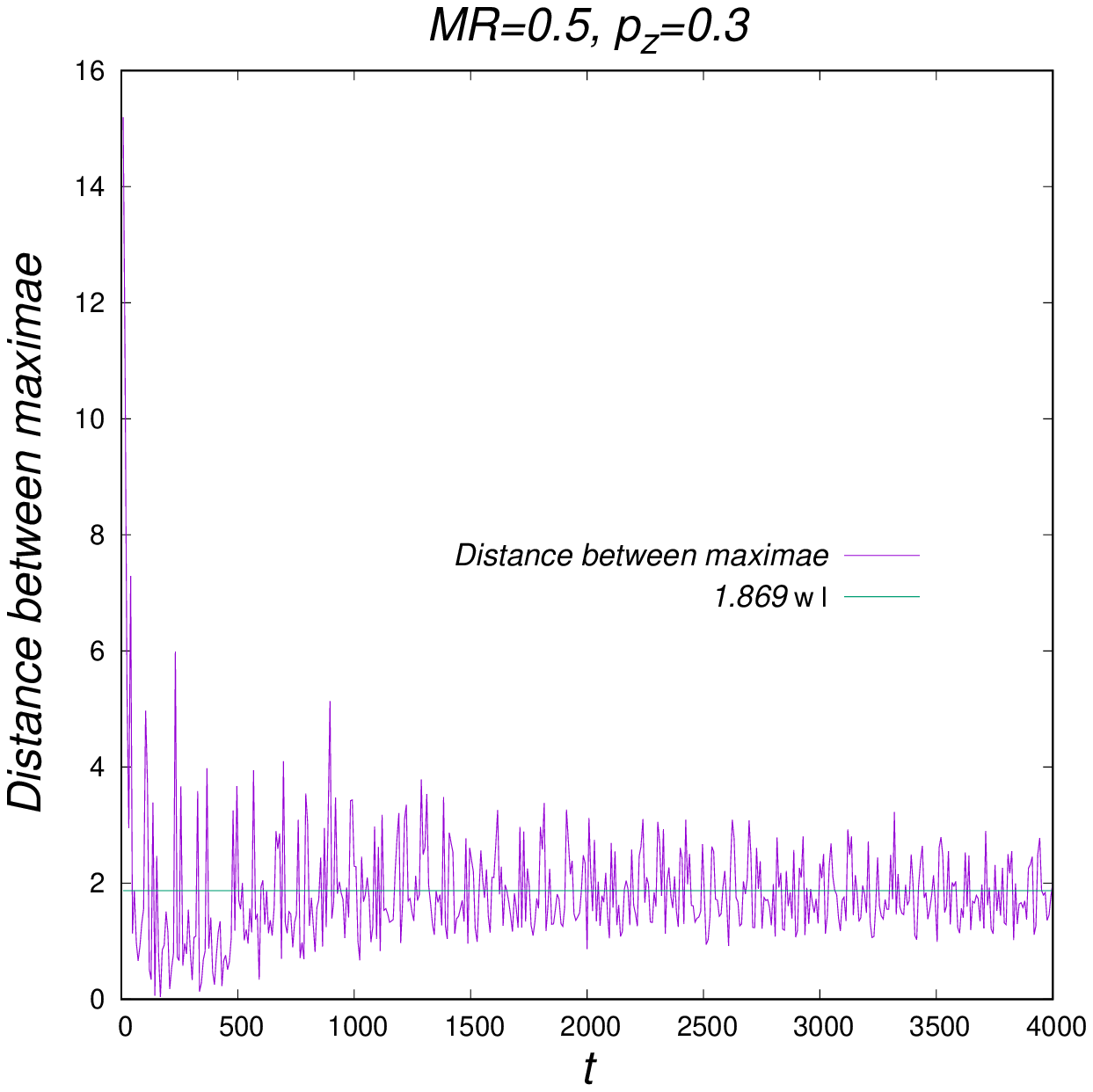} 
\caption{\label{fig:centers} Position of the maximum of $\rho_{1}$ in the first column, and $\rho_2$ in the second column, as function of time for various combinations of parameters. Notice that the maximum of $\rho_2$ oscillates around the maximum of $\rho_1$. In the third column we show the distance between the  maximums of $\rho_1$ and $\rho_2$, which oscillates around a fixed value, indicated with a green line, with decreasing amplitude. The behavior is generic for all combinations of parameters. }
\end{figure}

It is still possible to measure some properties of the distribution of the densities. We follow the maximum of $\rho_1$ and the maximum of $\rho_2$ as function of time in order to find whether there is a spatial relation in the distribution of the two states. The position of the maximum of $\rho_1$ along the $z-$axis is shown in the first column of Figure \ref{fig:centers}, which oscillates around the origin at $z=0$ with a clear dominant mode. Density $\rho_2$ has two local maximums, one to the right and one to the left of the maximum of $\rho_1$; we track the  position of the global maximum of $\rho_2$ as function of time and show it in the middle column of Fig. \ref{fig:centers}. The behavior of this maximum illustrates that the amplitude of $\rho_2$ is sometimes bigger on the right and some other on the left from the maximum of $\rho_1$ as barely seen in the snapshots of Fig. \ref{fig:mr0_9_pz0_2_snaps}. In the third column of Fig. \ref{fig:centers} we show the absolute value of the distance between the position of the global maximums of $\rho_1$ and $\rho_2$ as function of time, which asymptotically oscillates around a constant value.

\begin{figure}[htp]
\includegraphics[width=4.25cm]{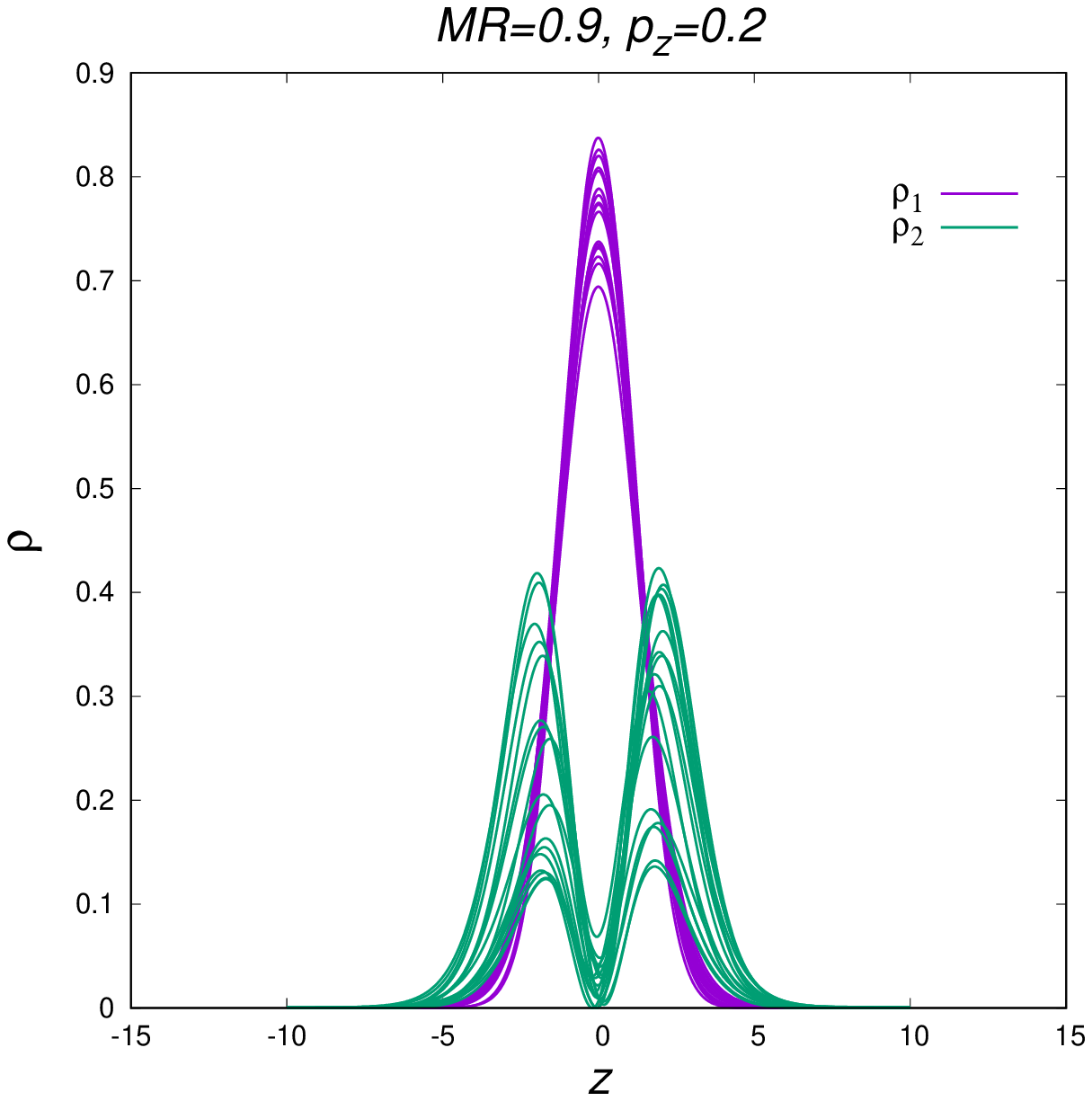} 
\includegraphics[width=4.25cm]{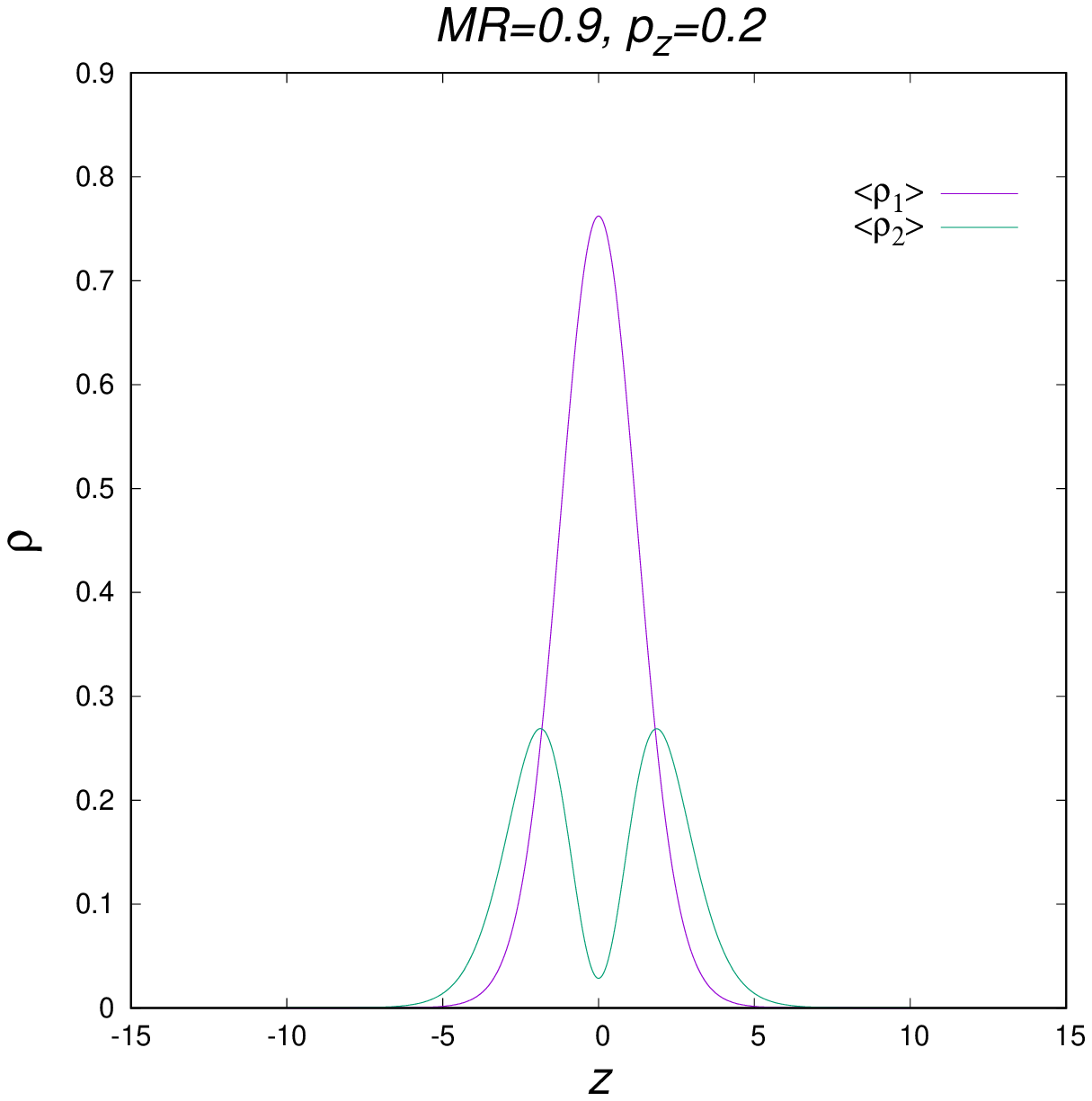} 
\caption{\label{fig:snapssubtracted} (Left) Snapshots of $\rho_1$ and $\rho_2$ for the case $MR=0.9$ and $p_{1z}=0.2$, in the reference system where the maximum of $\rho_1$ is at the origin. These snapshots indicate that states are in permanent motion. (Right) Densities $\rho_1$ and $\rho_2$ averaged in time, where some resemblance with equilibrium configurations from Figure \ref{fig:equilibrium} can be observed.}
\end{figure}

In order to study the motion of the system with respect to the center of the configuration, we subtract the oscillating motion plotted in the left column of Fig. \ref{fig:centers}. The dynamics of the two states for a generic case is illustrated in Fig. \ref{fig:snapssubtracted} and shows that $\rho_1$ and $\rho_2$ oscillate in time. This motion by itself indicates that the density profiles do not tend to a stationary state within the time window of the simulations.

Nevertheless, it is possible to investigate whether in average the distribution of the two states has a particular profile. We thus calculate the average in time of the snapshots in the simulation and the result appears in the right panel of Fig. \ref{fig:snapssubtracted}. There is a resemblance with equilibrium configurations from Fig. \ref{fig:equilibrium}, the density distributions are very similar in shape and distribution. Also important differences arise, for example, the density $\rho_2$ is not zero at the origin in average, as required for the state $l=1$.

\begin{figure}[htp]
\includegraphics[width=4.25cm]{mr0_9_pz0_2_averaged.eps} 
\includegraphics[width=4.25cm]{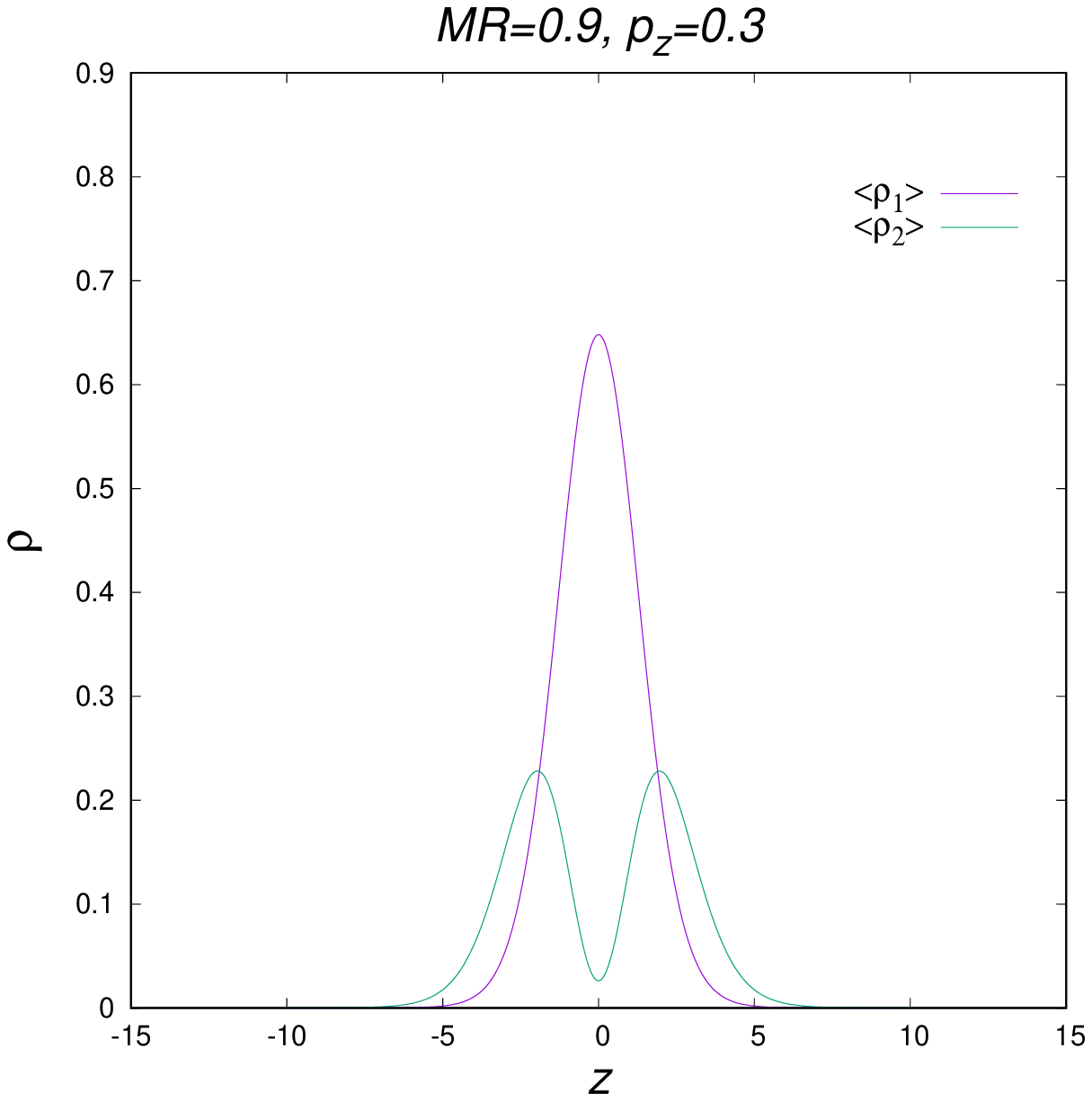} 
\includegraphics[width=4.25cm]{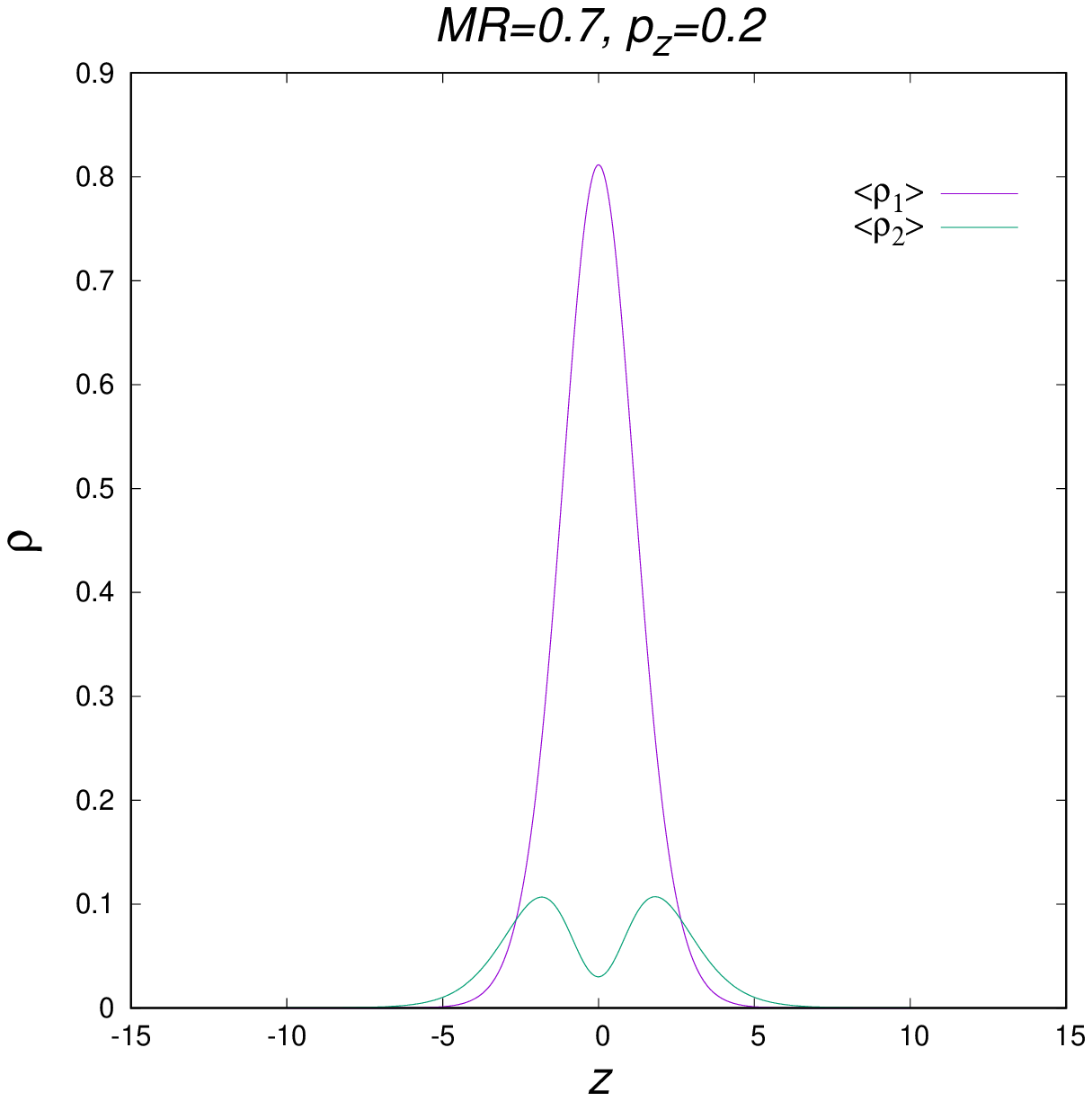} 
\includegraphics[width=4.25cm]{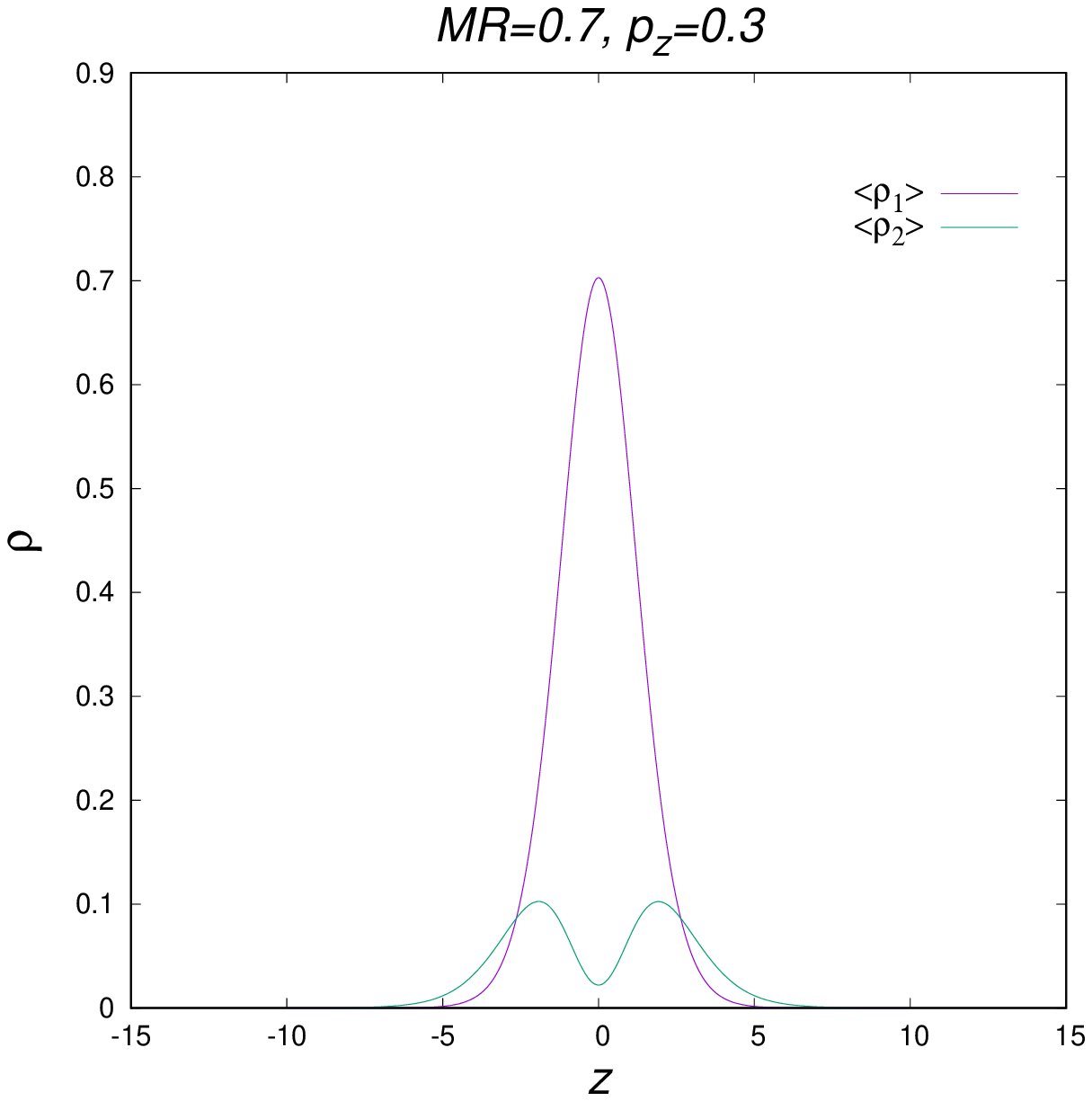} \\
\includegraphics[width=4.25cm]{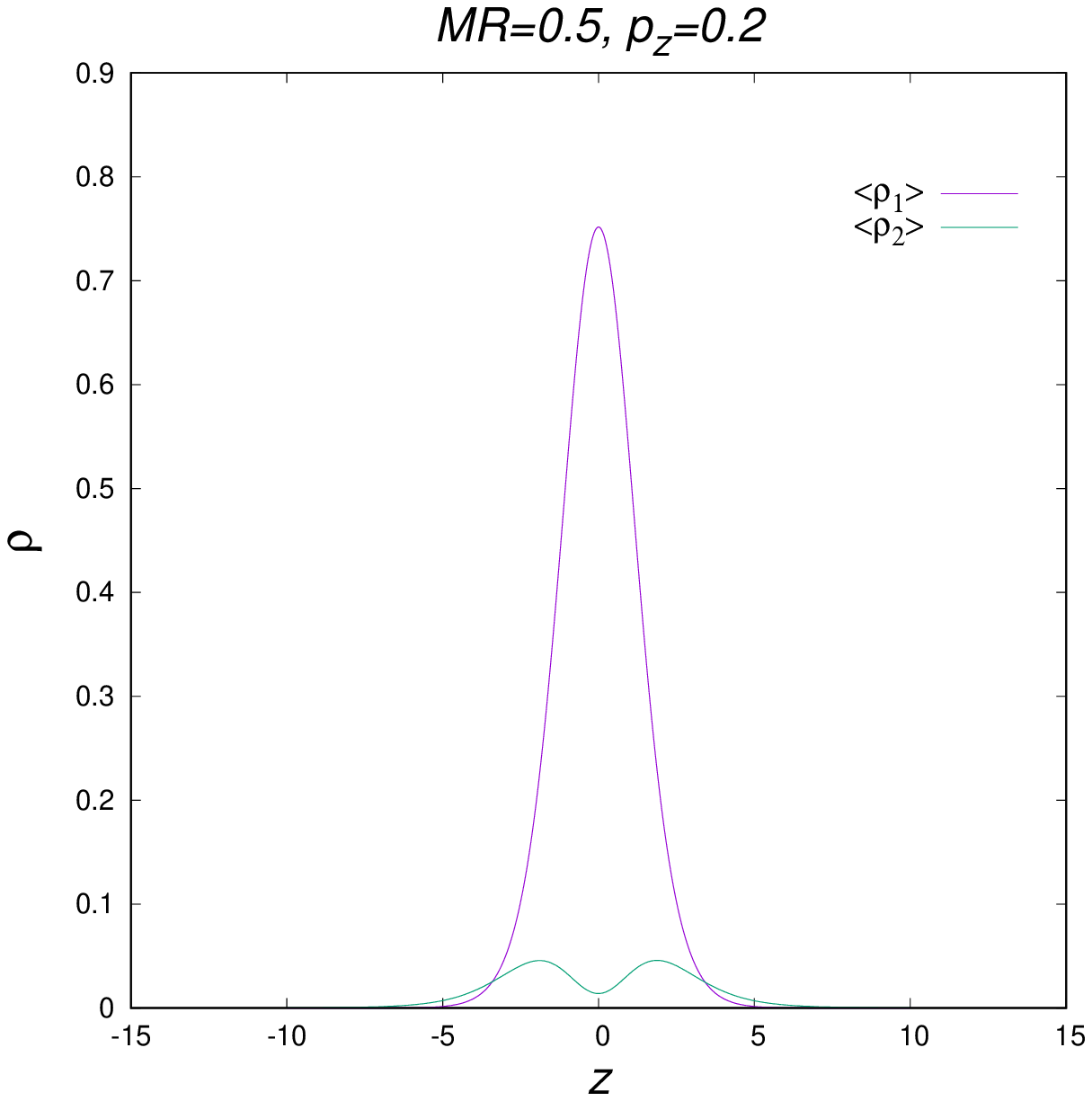} 
\includegraphics[width=4.25cm]{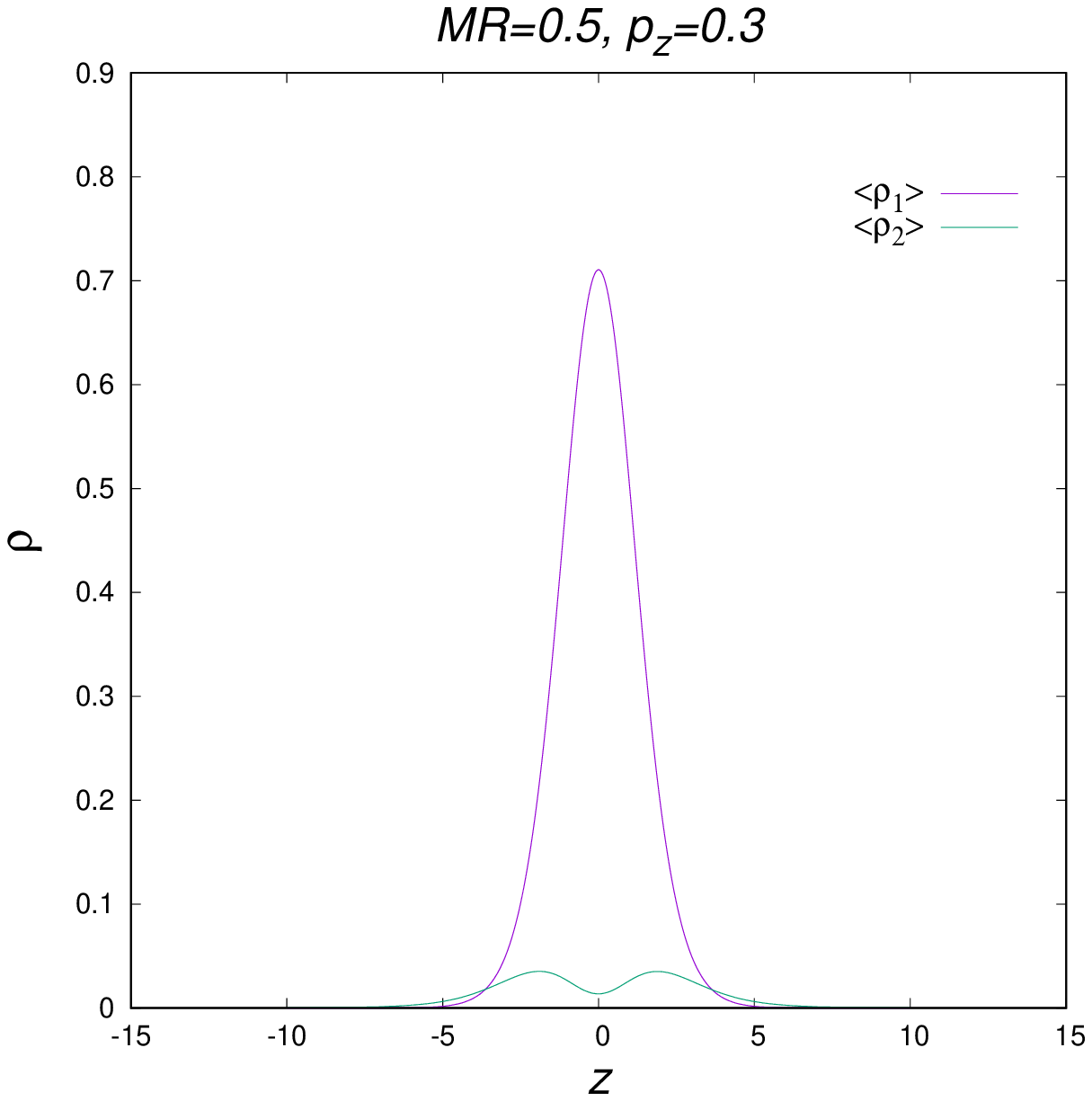} 
\caption{\label{fig:snapssubtractedALL} Time averaged density profile for various  combinations of parameters,  showing how generic the behavior is.}
\end{figure}

We apply this procedure to all the combinations of our parameter space and show the time-averaged density profiles $\langle \rho_1 \rangle$ and $\langle \rho_2 \rangle$ in Fig. \ref{fig:snapssubtractedALL}.

\section{Conclusions}
\label{sec:conclusions}

We have explored a formation mechanism of $(1,0,0)+(2,1,0)$ states of the Schr\"odinger-Poisson system, consisting on the head-on collision of two ground state equilibrium configurations in orthogonal states, coupled through Poisson equation. The findings are that during the collision, the configuration with the smaller mass pinches off the more massive one and distributes its density along the axis of collision with a morphology similar to that of the $(2,1,0)$ state.

According to our simulations, the resulting configuration does not stabilize within the time domain of the simulations, showing an apparently never-ending motion redistributing the density of each state, however keeping the morphology of the mixed state. Thus, strictly speaking, the configuration does not approach asymptotically toward a mixed-state equilibrium configuration.

Nevertheless, globally the configuration oscillates around a virialized state, and the energy and mass of each state stabilize around a constant value after the emission of exceeding matter of each state, a process consistent with the gravitational cooling relaxation process \cite{SeidelSuen1990,GuzmanUrena2006}.
Motivated by these global properties of the system, we explored the average behavior in time of the two states which reflects morphological and quantitative properties similar to those of equilibrium configurations.
In summary, the results indicate only a proof of principle of a formation mechanism of multistate configurations.

In an astrophysical context, this result can be of interest within the context of ultralight bosonic dark matter, which apparently offers an explanation of the anisotropic distribution of satellite galaxies \cite{jordi2021}. In a general context our results indicate a potential formation mechanism of multistate configurations of gravitational atoms. 


\section*{Acknowledgments}
This research is supported by grants CIC-UMSNH-4.9 and CONACyT Ciencias de Frontera Grant No. Sinergias/304001. The runs were carried out in the Big Mamma cluster at Laboratorio de Inteligencia Artificial y Superc\'omputo, IFM-UMSNH.


\bibliography{Paper_Arxiv}

\end{document}